\documentclass[showpacs,amssymb,twocolumn,floatfix,pre,aps]{revtex4-1}

\usepackage{amssymb,amsfonts,amsmath,bm}
\usepackage{graphics,graphicx}
\usepackage{xcolor}

\begin{document}

\title{A strong-coupling theory for polarizable symmetrically charged walls with counterions only}

\author{Ladislav \v{S}amaj}
\affiliation{Institute of Physics, Slovak Academy of Sciences,
D\'ubravsk\'a cesta 9, 84511 Bratislava, Slovakia}

\author{Alexandre P. dos Santos}
\affiliation{Instituto de F\'{i}sica, Universidade Federal do Rio Grande 
do Sul, CP 15051, CEP 91501-970 Porto Alegre, RS, Brazil}

\author{Emmanuel Trizac}
\affiliation{
Universit\'e Paris-Saclay, CNRS, LPTMS, 91405 Orsay, France 
}
\affiliation{
\'Ecole Normale Sup\'erieure de Lyon, 69342 Lyon, France}

\date{\today} 

\begin{abstract}
A pair of parallel polarizable planar walls at distance $d$ is considered.
The walls are symmetrically charged with a uniform surface charge density,
neutralized by mobile point counterions moving between them.
The case of repulsive particle images is studied in the strong-coupling (SC)
regime.
Of interest is the dependence of the effective inter-wall interaction
(pressure), mediated by the mobile counterions, as a function of the distance $d$.
It is shown that previous virial SC single-particle theories
work well at small $d$ when the dielectric jump is small; for intermediate
and large dielectric jumps they are inadequate even in the SC region. 
Here, we propose a Wigner-type SC theory based on harmonic
deviations of particles from their ground-state monolayer or bilayer
Wigner structures formed inside the space between the dielectric walls.
Our Monte-Carlo simulations are in very good agreement with
the Wigner SC predictions, even down to moderate coupling constants ($\Xi\geq 10$).
\end{abstract}

\maketitle

\renewcommand{\theequation}{1.\arabic{equation}}
\setcounter{equation}{0}

\section{Introduction} \label{sec:introduction}
Experiments with mesoscopic colloids are often carried out in water,
a polar solvent that strips mobile charges from colloid surfaces.
As a result, colloidal surfaces become charged with a modulated
surface charge density; mobile particles of opposite charge are known as
``counterions''.
Although ions in liquids have both signs, it is possible to experimentally
obtain deionized (salt-free) suspensions with ``no coions'', i.e.
``with counterions only'' \cite{Raspaud00,Palberg04,Brunner04}.
Counterions near a charged surface form an electric double layer
\cite{Attard96,Levin02,Messina09}.
The effective interaction between two similarly charged electric double layers, mediated by
counterions, is of particular experimental and theoretical interest
\cite{BenTal95,Jonsson99,Hansen00,Belloni00,Grosberg02}.
To simplify the theoretical study of complex Coulomb systems,
the curved surface of a large colloid is approximated by a planar surface
and the modulated charge density on the colloid surface by a uniform density.
As a first step in the theoretical treatment of electric double layers, the dielectric jump
between the wall and the solvent is usually ignored to avoid complicated
manipulations with infinite arrays of dielectric images. 

Coulomb particle systems are usually treated in the framework of
the Poisson-Boltzmann mean-field theory or its linearized version,
the Debye-H\"{u}ckel theory, which are valid in the high-temperature
(weak-coupling, WC) limit \cite{Gouy10,Chapman13,Chan76,Derjaguin87}.
Like-charged macroions always repel each other in the high-temperature
regime.

The opposite low-temperature (strong-coupling, SC) regime is more interesting
due to the presence of counterintuitive phenomena such as overcharging
or like-charge attraction, see reviews
\cite{Levin02,Messina09,Grosberg02,Naji05,Trizac12}.
The first evidence of the attraction of like-charged macromolecules was
observed experimentally
\cite{Khan85,Kjellander88,Bloomfield91,Kekicheff93,Dubois98}
as well as by computer simulations \cite{Guldbrand84,Kjellander84,Gronbech97}.
For the SC regime of electric double layers without dielectric jump, various types of theoretical
approaches have been proposed. 
The SC approach based on the virial expansion of the field-theoretical
representation of Coulomb systems \cite{Moreira00,Moreira01,Netz01},
designated as VSC theory, implies the leading term for the density profile
which corresponds to the picture of a single particle in
the electric potential of charged wall(s).
The validity of this leading term has been confirmed by Monte Carlo (MC)
simulations \cite{Moreira00,Moreira01,Dean09}.
The leading single-particle VSC theory has been extended to curved (spherical
and cylindrical) wall geometries \cite{Naji05}, image charge effects
\cite{Kanduc07,Jho08}, asymmetrically charged plates \cite{Kanduc08},
the presence of salt \cite{Kanduc10}, etc.
Next correction orders of the VSC approach in inverse powers of
the coupling constant have correct functional forms in space,
but wrong prefactors.
Another series of SC approaches, designated as Wigner Strong-Coupling theories (WSC), is based on
the formation of classical two-dimensional Wigner crystals of counterions on
wall surfaces at zero temperature \cite{Grosberg02,Levin99,Shklovskii99,dosSantos09,dosSantos10}.
Taking into account the harmonic deviations of counterions from their
ground-state Wigner positions in Refs. \cite{Samaj11a,Samaj11b}
correctly reproduces the leading single-particle theory of the VSC approach.
Moreover, the first WSC correction term for the counterion density
is in very good agreement with MC data over a wide range of strong and
intermediate Coulombic couplings.
To link the WSC theory to the fluid phase, which originates from the
crystal phase via a phase transition at very low temperatures,
the Wigner lattice structure was replaced by a correlation
hole in Ref. \cite{Palaia18}.
It was shown in Ref. \cite{Palaia22} that the relevant physics for
the like-charge attraction of the walls at intermediate and large distances
comes from the ground-state pressure. 

Ignoring the dielectric jump between the wall and the solvent
is not an adequate approximation for experimental situations.
In general, the dielectric constant of the interior of a colloid is much
smaller than that of the solvent (water), which implies repulsive dielectric
images of particle charges and surface charge densities.
The single-particle VSC route to account for repulsive image charges
was developed in \cite{Kanduc07,Jho08}.
As is shown in the present paper, the VSC theory is applicable only
to small distances and small values of the dielectric jump.
In the limit of large distances the (dimensionless) VSC pressure
goes to a nonzero constant and not to 0 as it should be.
For large values of the dielectric jump, the single-particle VSC theory
fails to reproduce MC data even in the region of strong couplings
which should provide its ``rigorous'' starting point.

The first attempt to apply the WSC theory to the geometry of a single planar
electric double layer with a dielectric jump was proposed in Ref. \cite{Samaj12}.
In the case of repulsive images, the ground state of counterions
corresponds to a stable 2D Wigner hexagonal crystal at some distance 
$l>0$ from the wall, the value of which is given by the balance
of the attractive force between the Wigner monolayer and the uniform surface
charge density on the one hand, and the repulsive force between the Wigner crystal and
its dielectric image on the other hand.
Analogously to the WSC theory developed in \cite{Samaj11a,Samaj11b},
taking into account harmonic deviations of the counterions from their
ground-state Wigner positions leads to realistic density profiles with
a maximum peak at distance $l$ from the wall, over a large range of strong
and intermediate couplings.
A simplified liquid-like description, replacing the Wigner structure
with the nearest neighbors of a reference particle, has been proposed for
intermediate and low couplings in Ref. \cite{Samaj16}.
The agreement with the MC data improves with increasing coupling
constant, as it should be.
Since the interactions of counterions within a collective ground state
of Wigner type are crucial in the strong coupling limit,
the single-particle scenario is not adequate for description
of the single electric double layer with dielectric jump in the SC regime.

In this paper, we extend the WSC approach to the geometry of two parallel
polarizable electric double layers charged symmetrically with same surface charge densities.
The strategy is similar as in previous applications of the WSC method,
i.e., starting from the ground state of counterions and considering
the harmonic deviations of particles from their ground-state positions.
The most significant complication is to find and to manipulate
the non-trivial ground state of counterions for given values of
model's parameters.
In general, the counterions form a bilayer Wigner crystal.
For planar interfaces without a dielectric jump, the charges condense
evenly on the opposite walls due to Earnshaw's theorem \cite{Earnshaw}. 
As the distance between the walls $d$ increases, the counterions take on
five distinct phases
\cite{Falko94,Esfarjani95,Goldoni96,Messina03,Lobaskin07,Oguz09}. 
In the case of repulsive image charges, Earnshaw's theorem no longer holds
and counterions can occupy the space between the walls.
There are five distinct phases similar to those observed in the previous
case of zero dielectric jump, but the bilayer structure of the counterions
lies inside the space region between the walls \cite{Samaj12c}, rather than at the walls.

Mobile counterions between the walls mediate the effective force
between electric double layers, as expressed by the pressure.
This quantity is usually calculated using the so-called contact value theorem,
which connects the pressure to the density profile and surface charge
density.
For planar interfaces with no dielectric jump, the pressure depends
only on the particle number density at the interface
and the surface charge density
\cite{Carnie81,Henderson78,Henderson79,Blum81,Wennerstrom82}.
The extension of the contact value theorem to planar electric double layers
with dielectric discontinuity \cite{Carnie81b,Jancovici82} leads to
the presence of two-point particle correlation functions in the formalism,
while the field-theoretical approach induces an additional Casimir term
\cite{Dean03}. 
This is why we calculate the pressure in an alternative although standard way within
the canonical formalism as (minus) the derivative of the free energy
with respect to the distance $d$ between the walls.

There is a wealth of simulation methods developed to account for the present
setup, with a dielectric contrast
\cite{Tyagi10,Jadhao12,Zwanikken13,Gan15,dosSantos15}.
In this work the method based on periodic Green functions is applied
\cite{dosSantos17}.
The present MC data as well as the explicit results of the WSC theory
show that for any value of the dielectric jump, the pressure in the SC regime
is close to its ground-state counterpart, with a well-specified interval
of inter-wall distances in which the pressure is negative (attractive).
Moreover, the WSC pressure is in perfect agreement with MC data down to
weak couplings $\Xi\simeq 10$, where the pressure is positive for any distance
between the walls.
The pressure calculated by the VSC theory fails to reproduce MC data
in the region of intermediate and large distances between the walls and
for intermediate and large values of the dielectric jump, even in
the SC limit.

The paper is structured as follows.
In Sec. \ref{sec:groundstate} we recapitulate the Wigner ground-state
monolayer and bilayer structures for the case of repulsive images.
Useful formulas for the ground-state pressure are derived in
Appendix \ref{appA} and \ref{appB} for relevant structures (phases referred below as structures I, II and III).
A short review of the VSC theory is given in Sec. \ref{sec:VSC}
with remarks concerning the small- and large-distance behavior of
the VSC pressure.
Our WSC approach, taking into account harmonic deviations of counterions
from their Wigner structure, is developed in Sec. \ref{sec:SC}.
Details of the MC simulations are given in Sec. \ref{sec:MC}.
A comparison of the VSC and WSC theories with data of MC simulations
is presented in Sec. \ref{sec:results}.
Concluding remarks are given in Sec. \ref{sec:conclusion}.

\renewcommand{\theequation}{2.\arabic{equation}}
\setcounter{equation}{0}
   
\section{Wigner monolayer and bilayer ground states} \label{sec:groundstate}

\begin{figure}[]
\begin{center}
\includegraphics[clip,width=0.4\textwidth]{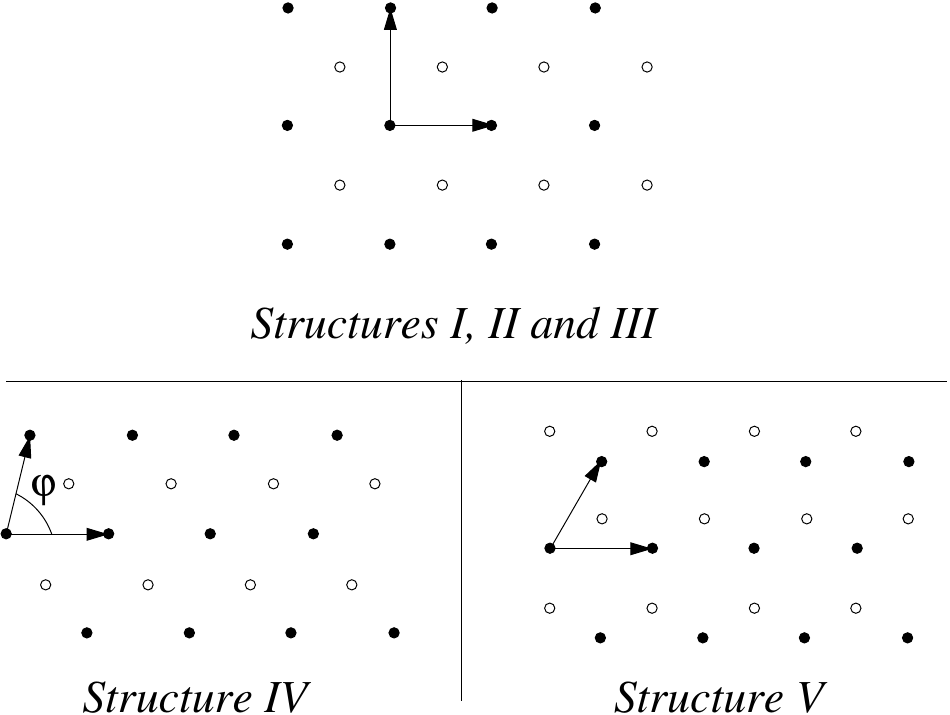}
\caption{Schematic plots of the ground states of counterions forming
the bilayer with sub-lattices $A$ (filled symbols, for ions lying all in the same plane) and $B$ (open symbols, for ions all lying in the same plane, in general different from that of sub-lattice $A$)
when the wall distance $d$ increases.
The lattice vectors are depicted by arrows, the ratio of their amplitudes
defines the aspect ratio $\delta$. 
Phases/structures I-V are described in the text.}
\label{fig1}
\end{center}
\end{figure}

Let us first review ground states for planar interfaces with no dielectric
jump between the wall and the solvent in which the particles are immersed.
For this case, Earnshaw's theorem \cite{Earnshaw} specifies that the particles
constrained to move in a domain condense onto the boundaries of this domain. 
In the geometry of one wall with a uniform surface charge density, the particles
stick to the wall surface and form a hexagonal (equilateral triangle) lattice.
We mention that real systems do not feature uniformly charged surfaces, but exhibit some degree 
of heterogeneity in the charge distribution. The very nature of the structure of counterions in their vicinity is thus a complex problem. Yet, we expect that the present approach captures a key feature of such structures in strong coupling, with the notion of correlation hole: there is always a region void of counterions in the immediate vicinity of a given counterion. It is this correlation hole that is at the root of the like-charge attraction (negative pressures) reported here. Besides, due to the repulsive nature of the image charges, counter-ions are expelled from the immediate vicinity of the charged interface, which smoothes the effect of surface charge heteronegeities.

In the case of two parallel planar symmetrically charged interfaces,  
the charges condense on the opposite wall surfaces and create on them
specific staggered lattice structures, that we denote $A$ and $B$.
As the distance between the walls $d$ increases, the counterions arrange in
five distinct structures (or phases)
\cite{Falko94,Esfarjani95,Goldoni96,Messina03,Lobaskin07,Oguz09}. 
Schematic plots of the possible bilayers composed of the $A$ (filled symbols)
and $B$ (open symbols) sub-lattices are pictured in Fig. \ref{fig1}.
For $d=0$, a staggered hexagonal Wigner bilayer with the aspect ratio of
the lattice vectors $\delta=\sqrt{3}$ (phase I) provides the
true minimum of the energy (i.e., it is stable).
Phase I becomes unstable for an arbitrary small $d>0$
\cite{Oguz09,Samaj12a,Samaj12b}. 
In the opposite $d\to\infty$ limit, the two layers decouple and 
a hexagonal Wigner crystal is formed on each wall surface and 
a staggered configuration of these two Wigner crystals minimizes
the inter-layer repulsion (phase V).
For intermediate values of $d$, a staggered rectangular lattice
[phase II with $\delta\in (1,\sqrt{3})$], a staggered square lattice
(phase III with $\delta=1$) and a staggered rhombic lattice
(phase IV with a varying angle $\varphi$) \cite{Goldoni96}
take place as $d$ increases, see Fig. \ref{fig1}. 
The transitions between phases (II,III) and (III,IV) are continuous
(of second order) with mean-field critical indices \cite{Samaj12a,Samaj12b},
while the transition between phases (IV,V) is discontinuous
(of first order).

\begin{figure}[]
\begin{center}
\includegraphics[clip,width=0.4\textwidth]{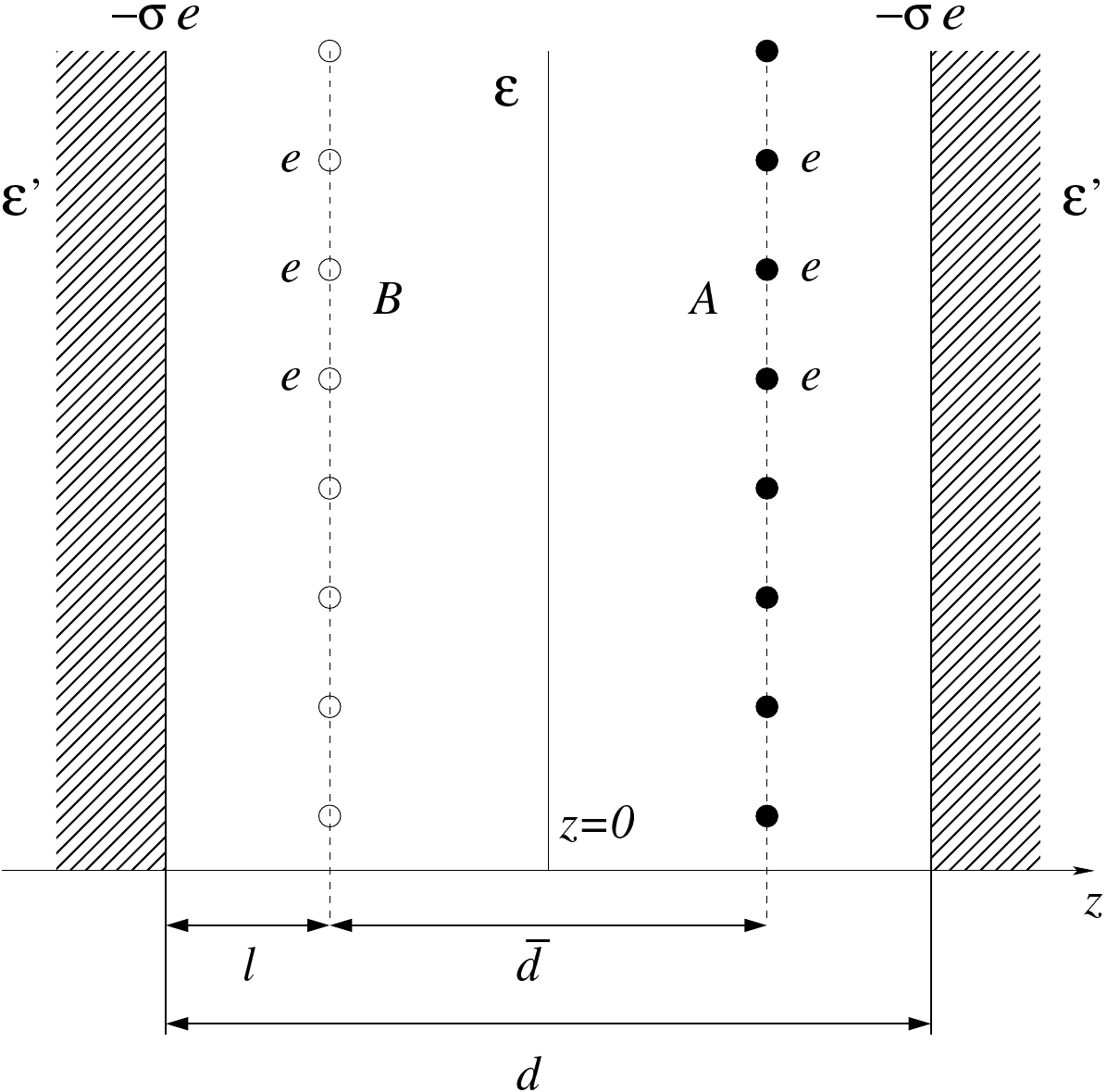}
\caption{The geometry of two parallel polarizable planar walls at distance $d$
along the $z$-axis, charged symmetrically by the surface charge density
$-\sigma e$.
In the ground state, counterions of charge $q e$ form a Wigner bilayer, i.e.,
two sub-lattices $A$ and $B$ in the $(x,y)$-plane at distance $\bar{d}$.
The distance of sub-lattices $A$ and $B$ to their nearest wall
is given by $l=(d-\bar{d})/2$.}
\label{fig2}
\end{center}
\end{figure}

In the case of dielectric inhomogeneity with {\em repulsive} image charges,
the Earnshaw theorem no longer holds and the counterions are allowed to
occupy also the interior space of the domain they are constrained to.
In the geometry of one wall with a uniform surface charge density,
the counterions form a hexagonal lattice at a distance from the wall
determined by the balance of the particle attraction to the surface charge
density and the particle-image repulsion \cite{Samaj12,Samaj16}.
The modulation of the surface charge density is not important at this regime
and the surface charge will look uniform.
In the case of two parallel planar symmetrically charged interfaces, 
the counterions form a monolayer structure in the center between the walls
of the symmetric bilayer structure pictured in Fig. \ref{fig2}.
Here, $\varepsilon'$ and $\varepsilon$ are the static dielectric constants
of the walls and of the solvent in which the particles are immersed,
respectively.
The dielectric jump between the walls and the solvent is defined by
the parameter
\begin{equation}
\Delta = \frac{\varepsilon - \varepsilon'}{\varepsilon + \varepsilon'} .
\end{equation}  
The considered case of repulsive images corresponds to the interval
$\Delta\in (0,1]$.
In real experiments, the wall mimics the colloid's interior with
$\varepsilon'\leq 10$ and $\varepsilon\simeq 80$ for water solvent,
so that $\Delta$ is close to $1$.
The cases with $\Delta<0$ (attractive image charges) would lead to
a divergence in the coulomb interaction, because the ions are point-like in the present model.
In a future study, the theory will be extended to consider hydrated ions near polarizable
surfaces with $\Delta<0$.  It is important to mention that small hard cores do not play an essential role
in the SC limit, but large hard cores which interfere with Wigner structure are important \cite{Samaj20}.
The counterions lie on the sub-lattices $A$ and $B$, their mutual distance
$\bar{d}$ being less than the distance $d$ between the walls.

In analogy with the previous case of zero dielectric jump, there exist
five distinct phases as the distance between the walls $d$ increases
\cite{Samaj12c}, see Fig. \ref{fig1}.
Phase I, i.e. the hexagonal monolayer localized in the center between
the walls ($\bar{d}=0$) composed of two interwoven rectangular lattices with
the aspect ratio $\delta=\sqrt{3}$, occurs in the region of small $d$.
Phase I transforms into phase II via a continuous transition at a
distance $d_{\rm I\to II}$.
Phase II corresponds to a bilayer Wigner structure ($\bar{d}>0$),
namely a staggered rectangular lattice with the aspect ratio $\delta$
which decreases continuously from $\sqrt{3}$ at distance $d_{\rm I\to II}$
to the square lattice value $1$ at distance $d_{\rm II\to III}$; phase III is
a staggered square lattice.
Phases IV and V correspond to staggered rhombic and hexagonal bilayers,
respectively.

To determine the interval of distances $d$ in which a given phase dominates
over other phases, it is necessary to express explicitly the corresponding
ground-state energy in terms of phase's parameters.
In the following we give a brief overview of the formulas from Ref.
\cite{Samaj12c} that are needed for the analytic expression of the
ground-state energy of a bilayer structure depicted in Fig. \ref{fig2}.
Each point in the three-dimensional Euclidean space is represented by
Cartesian coordinates ${\bf r}=({\bf R},z)$, where ${\bf R}=(x,y)$
and the $z$-axis is perpendicular to the pair of parallel wall surfaces.
The wall surfaces, localized at $z=\pm d/2$, are charged uniformly
by the surface charge density $-\sigma e$ where $e$
is the elementary charge.
The $N$ (even $N$) point-like particles of charge $e$ form two Wigner
sublayers $A$ and $B$ localized at $z=\bar{d}/2$ and $z=-\bar{d}/2$,
respectively, each containing the same number
$j=1,2,\ldots,N/2$ and $j=N/2+1,N/2+2,\ldots,N$ charges.
The sites of sub-lattices $A$ and $B$ will be denoted as $\{ {\bf R}_j^A\}$
and $\{ {\bf R}_j^B\}$.
Denoting by $S$ the surface of either of the walls, the overall
electroneutrality of the system corresponds to the condition
\begin{equation} \label{electroneutrality}
N = 2\sigma S .
\end{equation}  

In Gauss units, the direct Coulomb potential between the points
${\bf r}=({\bf R},z)$ and ${\bf r}'=({\bf R}',z')$ in between the two walls
is given by
\begin{equation} \label{Coulomb}
u_0({\bf r},{\bf r}') = \frac{1}{\varepsilon}
\frac{1}{\sqrt{\vert {\bf R}-{\bf R}'\vert^2 + (z-z')^2}} .
\end{equation}
In the presence of the dielectric jump between the walls and the solvent,
each particle localized at ${\bf r}=({\bf R},z)$ has an infinite number
of electrostatic images with the same component ${\bf R}$ \cite{Jackson}.
Taking as a first reflection the one with respect to the wall surface at
$z=+d/2$, an infinite sequence of images with charges $\Delta^n$
is generated at positions $(-1)^{n+1} (n d - z)$ $(n=1,2,\ldots)$.
Taking as a first reflection the one with respect to the wall surface at
$z=-d/2$, an infinite sequence of images with charges $\Delta^n$
is generated at positions $(-1)^n (n d + z)$ $(n=1,2,\ldots)$.
The image-charge Coulomb potential between the points ${\bf r}=({\bf R},z)$
and ${\bf r}'=({\bf R}',z')$ in between the two walls then reads as
\begin{eqnarray} 
u_{\rm im}({\bf r},{\bf r}') & \equiv & u_{\rm im}({\bf R}-{\bf R}';z,z')
\nonumber \\ & = & \frac{1}{\varepsilon} \sum_{n=1}^{\infty}
\Bigg\{ \frac{\Delta^{2n}}{\sqrt{\vert {\bf R}-{\bf R}'\vert^2
+ (2nd+z-z')^2}} \nonumber \\
& & + \frac{\Delta^{2n}}{\sqrt{\vert {\bf R}-{\bf R}'\vert^2
+ (2nd-z+z')^2}} \nonumber \\
& & + \frac{\Delta^{2n-1}}{\sqrt{\vert {\bf R}-{\bf R}'\vert^2
+ \left[ (2n-1)d+z+z'\right]^2}} \nonumber \\
& & + \frac{\Delta^{2n-1}}{\sqrt{\vert {\bf R}-{\bf R}'\vert^2
+ \left[ (2n-1)d-z-z'\right]^2}} \Bigg\} . \nonumber \\
& & \label{image}
\end{eqnarray}
The total interaction energy of charged particles at positions
$\{ {\bf r}_j \}$ and their images is given by
\begin{equation}
E_{pp} = \frac{e^2}{2} \left[ \sum_{j\ne k} u_0({\bf r}_j,{\bf r}_k)
+ \sum_{j,k} u_{\rm im}({\bf r}_j,{\bf r}_k) \right] ,    
\end{equation}
where the $j=k$ term in the second sum on the rhs corresponds to
the interaction of the $j$th particle with its own image.
To express explicitly the energy per particle, let us take advantage of
equivalence of all particle positions and choose as a reference particle
the one at site 1 of sub-lattice $A$.
The energy per particle can be then written as
\begin{eqnarray}
\frac{E_{pp}}{N} & = & \frac{e^2}{2\varepsilon} \left[
\sum_{j\ne 1} \frac{1}{R_{1j}^{AA}} + \sum_j \frac{1}{
\sqrt{\left( R_{1j}^{AB}\right)^2 + \bar{d}^2}} \right] \nonumber \\
& & + \frac{e^2}{2} \Bigg[
u_{\rm im}\left( 0;\frac{\bar{d}}{2},\frac{\bar{d}}{2}\right) + \sum_{j\ne 1}
u_{\rm im}\left( R_{1j}^{AA};\frac{\bar{d}}{2},\frac{\bar{d}}{2}\right)
\nonumber \\ & & + \sum_j
u_{\rm im}\left( R_{1j}^{AB};\frac{\bar{d}}{2},-\frac{\bar{d}}{2}\right) \Bigg] ,  
\label{part-part}
\end{eqnarray}  
where we use the notation
$R_{1j}^{AA}\equiv \vert {\bf R}_1^A - {\bf R}_j^A\vert$ and
$R_{1j}^{AB}\equiv \vert {\bf R}_1^A - {\bf R}_j^B\vert$.

The lattice sums in (\ref{part-part}) contain separately infinite constants
due to the local system's non-neutrality; these infinite constants are zeroed
due to the overall neutrality of the system.
As shown in Ref. \cite{Samaj12c}, the lattice sums can be regularized by
a local neutralizing background of a uniform surface density of charge
opposite to that of the particles.
This procedure induces additional finite terms, namely
\begin{equation}
\frac{E_{pp}}{N} = \frac{E_{pp}^*}{N} - \frac{\pi\sigma e^2}{\varepsilon} \bar{d}
- \frac{4\pi\sigma e^2}{\varepsilon} \frac{\Delta}{(1-\Delta)^2} d ,  
\end{equation}
where the star means ``regularized by charge background''.
The energy of particles with the surface charge densities and their images
was found to be
\begin{equation}
\frac{E_{ps}}{N} = \frac{2\pi\sigma e^2}{\varepsilon}
\frac{(1+\Delta)^2}{(1-\Delta)^2} d .  
\end{equation}
The energy of the direct and image interactions between the surface charge
densities on the two plates is given by
\begin{equation}
\frac{E_{ss}}{N} = - \frac{\pi\sigma e^2}{\varepsilon}
\frac{(1+\Delta)^2}{(1-\Delta)^2} d .  
\end{equation}
Finally, the total energy $E=E_{pp}+E_{ps}+E_{ss}$ per particle
reads as \cite{Samaj12c}
\begin{equation} \label{gsenergy}
\frac{E}{N} = \frac{E_{pp}^*}{N} + \frac{\pi\sigma e^2}{\varepsilon}
(d-\bar{d}) .
\end{equation}  

In the letter \cite{Samaj12c}, we obtained the complete phase diagram of
the considered system by deriving exact expressions for the ground-state
energy of phases I-V as integrals over certain products of the Jacobi theta
functions with zero argument \cite{Gradshteyn} by using a technique
developed in Ref. \cite{Samaj12b}.
Due to lack of space, we could not provide details of the calculations there.
Because these details are important in formulating the present WSC theory,
we demonstrate in appendices the calculation of the ground state energies for
phases I-III.
These phases cover a wide range of inter-wall spacings that is sufficient for
our purposes.

The ground-state energy of phase I is derived in Appendix \ref{appA},
see formula (\ref{E01}).
The pressure between the dielectric walls $P_0$ is given by
\begin{equation}
P_0(d) = - \frac{1}{S} \frac{\partial E_0}{\partial d}
= - 2 \sigma \frac{\partial (E_0/N)}{\partial d} ,  
\end{equation}  
where we used the electroneutrality condition (\ref{electroneutrality}).
Introducing the dimensionless pressure
\begin{equation} \label{dimpress}
\widetilde{P}_0(d) \equiv \frac{P_0(d)}{2\pi (e^2/\varepsilon)\sigma^2} , 
\end{equation}  
and the dimensionless distance $\eta = d\sqrt{\sigma}$,
we finally get
\begin{eqnarray}
\widetilde{P}_0(\eta) & = & - 1 - \frac{1}{\pi \eta^2} \ln(1-\Delta)
\nonumber \\ & & + \frac{2\eta}{\pi^{3/2}} \int_0^{\infty} {\rm d}t\, \sqrt{t}
\sum_{n=1}^{\infty} \Delta^n n^2 {\rm e}^{-(n\eta)^2 t} \nonumber \\ & & \times
\left[ \varphi_2(t,\sqrt{3}) + \varphi_3(t,\sqrt{3}) \right] , \label{P0eta}
\end{eqnarray}
where the functions $\varphi_2(t,\delta)$ and $\varphi_3(t,\delta)$
are expressed in terms of the Jacobi theta functions in (\ref{varphi2})
and (\ref{varphi3}), respectively.

The energy of phases II-III is derived in Appendix \ref{appB},
see formula (\ref{E02}).
For fixed model's parameters $\Delta$ and $\eta$, the energy per particle
$E/N$ depends on free variables $\bar{\eta}$ (defined as $\bar\eta = \bar d\sqrt\sigma$) and the aspect ratio $\delta$.
To obtain the ground-state energy, one has to minimize $E/N$ with respect
to these variables, i.e., to solve the system of equations
\begin{equation} \label{mincondition}
\frac{\partial E}{\partial \bar{\eta}} = 0 ,
\qquad \frac{\partial E}{\partial\delta}= 0  
\end{equation}  
for the $(\bar{\eta}^*,\delta^*)$ pair, such that the determinant
of the Hessian matrix
\begin{equation}
D(\bar{\eta},\delta) = \frac{\partial^2 E}{\partial \bar{\eta}^2}
\frac{\partial^2 E}{\partial \delta^2} -
\left( \frac{\partial^2 E}{\partial \bar{\eta} \partial \delta} \right)^2
\end{equation}
and $\partial^2 E/\partial \bar{\eta}^2$ are positive at the minimum
point $(\bar{\eta}^*,\delta^*)$.
In this way the characteristics $(\bar{\eta}^*,\delta^*)$ of the phases
I-III become the functions of model's parameters $(\Delta,\eta)$
and $E_0/N = E(\bar{\eta}^*,\delta^*)/N$, see figures 5 and 7
in Ref. \cite{Samaj12c}.
The pressure between the dielectric walls is given by
\begin{equation}
P_0(d) = - 2 \sigma \frac{{\rm d} (E_0/N)}{{\rm d} d}
= - 2 \sigma \frac{\partial (E_0/N)}{\partial d} ,  
\end{equation}  
where the total derivative with respect to $d$ can be interchanged by
the partial one because of the minimization conditions (\ref{mincondition}).
Because the form of the energy (\ref{E02}) is rather complicated,
the expression for the dimensionless pressure (\ref{dimpress})
is not written explicitly.

\renewcommand{\theequation}{3.\arabic{equation}}
\setcounter{equation}{0}

\section{Review of the virial SC theory} \label{sec:VSC}
At nonzero temperature $T$, or finite inverse temperature
$\beta=1/(k_{\rm B}T)$, there are two important length scales.
The Bjerrum length
\begin{equation} \label{Bjerrum}
\ell_{\rm B} = \frac{\beta e^2}{\varepsilon}
\end{equation}
is the distance at which thermal energy $k_{\rm B}T$ equals to
the Coulomb potential between two elementary charges $e$.
The Gouy-Chapman length \cite{Gouy10,Chapman13}
\begin{equation} \label{Chapman}
\mu = \frac{1}{2\pi \ell_{\rm B}\sigma} 
\end{equation}
is the distance from the wall charged by a uniform charge density
$-\sigma e$ at which an isolated counterion of charge $e$ has
the potential energy equal to $k_{\rm B}T$.
The dimensionless coupling parameter is defined as the ratio
\begin{equation}
\Xi = \frac{\ell_{\rm B}}{\mu} = 2\pi\ell_{\rm B}^2 \sigma .
\end{equation}  
Note that all parameters being fixed except the temperature $T$, the ground-state limit $T\to 0$ corresponds to $\Xi\to \infty$. 

The SC regime is defined as that of large $\Xi$. The
limit $\Xi\to\infty$ of the present model with dielectric
inhomogeneities was studied within the framework of the VSC theory in Refs.
\cite{Kanduc07,Jho08}.
In the VSC approach, the interactions of counterions with themselves
and their images are ignored.
A given particle is under the influence of the one-body electrostatic
potential induced by charged walls and their images as well as
an infinite set of self-images.
Distances along the perpendicular $z$ coordinate are usually measured
in units of $\mu$,
\begin{equation}
\tilde{z} \equiv \frac{z}{\mu} .  
\end{equation}
The walls in \cite{Jho08} were taken as membranes of finite thickness $b$,
the distance between the walls was denoted as $2a=d$.

According to the VSC method \cite{Jho08}, the dimensionless pressure acting
on each membrane
\begin{equation}
\widetilde{P}_{\rm VSC} \equiv \frac{\beta P_{\rm VSC}}{2\pi\ell_{\rm B}\sigma^2}
\end{equation}
is expressible as
\begin{equation}
\widetilde{P}_{\rm VSC}  = - \frac{\partial}{\partial \tilde{a}}
\widetilde{F}_{\rm VSC} ,   
\end{equation}
where $\widetilde{F}_{\rm VSC}$ is the dimensionless free energy per particle
defined by
\begin{widetext}
\begin{equation}
\widetilde{F}_{\rm VSC} = \tilde{a} - \ln\left[ \int_0^{\tilde{a}}
{\rm d}\tilde{z}\, \exp\left( -\Xi \int_0^{\infty} {\rm d}Q\, \psi(\tilde{z},Q)
\right) \right] ,  
\end{equation}  
where
\begin{equation} \label{psi}
\psi(\tilde{z},Q) =
\frac{\cosh(2Q\tilde{z})+\Delta_Q{\rm e}^{-2Q\tilde{a}}}{\Delta_Q^{-1}
{\rm e}^{2Q\tilde{a}} - \Delta_Q{\rm e}^{-2Q\tilde{a}}} , \qquad
\Delta_Q = \Delta \frac{1 - {\rm e}^{-2Q\tilde{b}}}{1
-\Delta^2 {\rm e}^{-2Q\tilde{b}}} .
\end{equation}  
In the present case of infinite membrane thickness $\tilde{b}\to\infty$
it holds that $\Delta_Q=\Delta$.
Since $\Delta {\rm e}^{-2Q\tilde{a}} < 1$, the denominator in $\psi$ (\ref{psi})
can be expanded in powers of $\Delta^2 {\rm e}^{-4Q\tilde{a}}$, with the result  
\begin{equation}
\psi(\tilde{z},Q) = \left[\frac{1}{2} \left(
{\rm e}^{2Q\tilde{z}}+{\rm e}^{-2Q\tilde{z}} \right)
+\Delta {\rm e}^{-2Q\tilde{a}} \right] \Delta {\rm e}^{-2Q\tilde{a}}
\sum_{n=0}^{\infty} \Delta^{2n} {\rm e}^{-4 n Q\tilde{a}} .
\end{equation}  
Since each term is an exponential of $Q$, the integral over $Q$ can be done
explicitly and after some simple algebra one gets
\begin{equation} \label{integralpsi}
\int_0^{\infty} {\rm d}Q\, \psi(\tilde{z},Q) =
\frac{1}{4} \sum_{n=1}^{\infty} \Delta^{2n-1} \left[
\frac{1}{(2n-1)\tilde{a}+\tilde{z}} + \frac{1}{(2n-1)\tilde{a}-\tilde{z}}
- \frac{2}{(2n-1)\tilde{a}} \right] - \frac{1}{2\tilde{a}} \ln(1-\Delta) .    
\end{equation}
As is shown in Appendix \ref{appA}, in addition to $\tilde a$ another natural
definition of the dimensionless distance between the walls is
$\eta=\sqrt{\sigma} d$, see Eq. (\ref{eta}).
Thus passing in (\ref{integralpsi}) from $\tilde{a}=\tilde{d}/2$ to
$\eta$ one obtains 
\begin{equation} \label{PVSC}
\widetilde{P}_{\rm VSC} = -1 - \frac{1}{\pi \eta^2} \ln(1-\Delta)
+ \sqrt{\frac{2}{\pi\Xi}} \frac{\partial}{\partial\eta}
\ln\left[ \int_0^{\eta/2} {\rm d}t\, {\rm e}^{-f_{\rm VSC}(t,\eta)} \right] ,
\end{equation}  
where
\begin{equation} \label{fVSC}
f_{\rm VSC}(t,\eta) = \frac{1}{2} \sqrt{\frac{\Xi}{2\pi}} \sum_{n=1}^{\infty}
\Delta^{2n-1} \left[ \frac{1}{(2n-1)\eta+2t} + \frac{1}{(2n-1)\eta-2t}
- \frac{2}{(2n-1)\eta} \right] .    
\end{equation}  
\end{widetext}

The VSC theory only takes into account one-body potentials acting on
a given particle which originate from the surface charges on the walls
and their images, as well from an infinite series of self-images
contained in $f_{\rm VSC}$.
Since the images and self-images are repulsive, the particle density
always has a maximum peak midway between the walls.
This property is adequate for small distances $d$ between the walls,
where monolayer phase I dominates at zero temperature, but the VSC
theory fails for larger distances, where bilayer phases with two
symmetric peaks of the particle density prevail. 

Fixing $\eta$ at some value, the dimensionless pressure (\ref{PVSC})
has a well defined SC $\Xi\to\infty$ limit.
The function $f_{\rm VSC}(t,\eta)$ (\ref{fVSC}) possesses the symmetry
$f_{\rm VSC}(t,\eta)=f_{\rm VSC}(-t,\eta)$ and its derivative with respect to $t$
is positive for all $t$.
Thus it can be expanded in powers of $t^2$ as follows
\begin{equation}
f_{\rm VSC}(t,\eta) = \sqrt{\Xi} \sum_{n=1}^{\infty} a_{2n}(\Delta,\eta) t^{2n} ,
\end{equation}
where the positive expansion coefficients read as
\begin{equation}
a_{2n}(\Delta,\eta) = \frac{1}{2\sqrt{2\pi}} \frac{\Delta}{\eta^{2n+1}}
\Phi(\Delta^2,2n+1,\tfrac{1}{2})   
\end{equation}  
with
\begin{equation}
\Phi(z,s,\alpha) = \sum_{j=0}^{\infty} \frac{z^j}{(j+\alpha)^s}
\end{equation}
being the Lerch transcendent \cite{Gradshteyn}.
After the substitution of variables $s=\Xi^{1/4}\sqrt{a_2}t$,
the integral in (\ref{PVSC}) becomes
\begin{equation}
\frac{1}{\Xi^{1/4}\sqrt{a_2}} \int_0^{\Xi^{1/4}\sqrt{a_2}\eta/2} {\rm d}s\,
\exp\left[ - s^2 - \frac{a_4}{\sqrt{\Xi}a_2^2} s^4
+ O\left( \frac{1}{\Xi} \right) \right] .
\end{equation}
By neglecting exponentially small terms in $\Xi$, the upper limit of
the integral can be moved to infinity.
Apart from the leading term $-s^2$, the exponential can be
expanded in powers of $s^2$ which implies convergent integrals.
Since $a_2\propto 1/\eta^3$ the leading prefactor of the integral
behaves like $\eta^{3/2}$, so
\begin{equation} 
\widetilde{P}_{\rm VSC} = -1 - \frac{1}{\pi \eta^2} \ln(1-\Delta)
+ \sqrt{\frac{2}{\pi\Xi}} \frac{3}{2\eta} + O\left( \frac{1}{\Xi} \right) .
\end{equation}  
Note that the thermal correction to the ground-state pressure 
\begin{equation} \label{PVSCgs}
\widetilde{P}_{\rm VSC}^{\rm gs} = -1 - \frac{1}{\pi \eta^2} \ln(1-\Delta) ,
\end{equation}  
which is of order $1/\eta$ when $\eta\to 0$, is subleading with respect to the
$1/\eta^2$ divergence of $\widetilde{P}_{\rm VSC}^{\rm gs}$.
We conclude that within the VSC theory, the divergence of the pressure
as $\eta\to 0$ is determined at any temperature by its ground-state
asymptotics.
According to the contact value theorem for the homogeneous dielectric
case $\Delta=0$, the dimensionless pressure does not diverge but goes
to the finite value $-1$ as $\eta\to 0$,
in agreement with the VSC formula (\ref{PVSCgs}).

Another interesting case is the large-distance limit $\eta\to\infty$.
Using the substitution of variables $t=s\eta/2$, the integral over $t$
on the rhs of (\ref{PVSC}) can be transformed to
\begin{equation} \label{gVSC}
\int_0^{\eta/2} {\rm d}t\, {\rm e}^{-f_{\rm VSC}(t,\eta)} =
\frac{\eta}{2} \int_0^1 {\rm d}s\, {\rm e}^{-g_{\rm VSC}(s)/\eta}  
\end{equation}  
where
\begin{eqnarray}
g_{\rm VSC}(s) & = & \frac{1}{2} \sqrt{\frac{\Xi}{2\pi}} \sum_{n=1}^{\infty}
\Delta^{2n-1} \left[ \frac{1}{(2n-1)+s} \right. \nonumber \\ & & \left.
+ \frac{1}{(2n-1)-s} - \frac{2}{(2n-1)} \right] .    
\end{eqnarray}  
The function $g_{\rm VSC}(s)$ is finite in the whole interval $s\in [0,1]$,
so all integrals $\int_0^1 {\rm d}s\, g_{\rm VSC}^n(s)$ $(n=1,2,\ldots)$
are finite as well.
Thus we can expand the exponential on the rhs of (\ref{gVSC}) in $1/\eta$,
with the result  
\begin{equation}
\frac{\eta}{2} \left[ 1 - \frac{1}{\eta} \int_0^1 {\rm d}s\, g_{\rm VSC}(s)
+ O\left( \frac{1}{\eta^2} \right) \right] .    
\end{equation}
Using this expansion in the formula for the dimensionless pressure
(\ref{PVSC}) yields
\begin{eqnarray}
\widetilde{P}_{\rm VSC}(\eta) & = & -1 - \frac{1}{\pi \eta^2} \ln(1-\Delta)
+ \sqrt{\frac{2}{\pi\Xi}} \left[ \frac{1}{\eta} \right. \nonumber \\ & & \left.
+ \frac{1}{\eta^2} \int_0^1 {\rm d}s\, g_{\rm VSC}(s) +
O\left( \frac{1}{\eta^3} \right) \right] .      
\end{eqnarray}  
Consequently,
\begin{equation}
\widetilde{P}_{\rm VSC}(\eta) \mathop{\sim}_{\eta\to\infty} -1 . 
\end{equation}  
This asymptotic result is in contradiction with the fact that the pressure
must vanish for an infinite distance between the walls: this deficiency of
the VSC theory hints at the fact that this approach cannot hold at arbitrary
large distances.
The Wigner type of theory below corrects this deficiency, in addition
to providing more accurate results.

We conclude this section by considering the particle density
$\rho(z)$ in the VSC approach.
It is proportional to the one-body Bolzmann factor
$\exp\left[-f_{\rm VSC}(\sqrt{\sigma}z,\eta)\right]$, normalized by
the electroneutrality condition (\ref{electroneutrality}) as 
\begin{equation}
\int_{-d/2}^{d/2} {\rm d}z\, \rho(z) = 2 \sigma . 
\end{equation}  
Explicitly,
\begin{equation} \label{rhoVSC}
\frac{\rho(z)}{2\pi\ell_{\rm B}\sigma^2} = \frac{1}{\sqrt{2\pi\Xi}}
\frac{\exp\left[-f_{\rm VSC}(\sqrt{\sigma}z,\eta)\right]}{
\int_0^{\eta/2} {\rm d}t\, \exp\left[-f_{\rm VSC}(t,\eta)\right]} .  
\end{equation}
The VSC density always exhibits a maximum at the center $z=0$ of
the wall distance.
This property is adequate if the model parameters correspond to phase I in
the ground state.
For phases II-V with double peaks, the VSC theory necessarily fails.

\renewcommand{\theequation}{4.\arabic{equation}}
\setcounter{equation}{0}

\section{Wigner SC theory} \label{sec:SC}
The Wigner-Strong Coupling theory aims at circumventing the shortcomings of the VSC approach, and stems from the fact that 
for $\Xi\to\infty$, the ground state situation is recovered. For a couple of charged plates 
without dielectric
inhomogeneity ($\Delta=0$), the WSC route was developed in Refs. \cite{Samaj11a,Samaj11b}.
The method is based on thermalization of harmonic deviations of counterions
from their Wigner ground-state positions at the surfaces of the walls,
in the direction perpendicular to surfaces of the walls.
Prefactors to harmonic deviations have relatively simple forms.
Here we proceed in the same way, but the task is technically much
more demanding because the Wigner structures are localized in between
the walls and particles induce an infinite series of electrostatic images
due to the presence of polarizable walls.
For simplicity, we shall restrict ourselves to phase I localized in
the middle between the walls with the aspect ratio $\delta=\sqrt{3}$.
Harmonic deviations along all three Euclidean directions contribute
to the pressure.
In what follows, we use techniques developed in Appendices \ref{appA} and
\ref{appB} to express prefactors to harmonic deviations in terms of
integrals over Jacobi theta functions.

Let us shift one particle, say the one on the reference site $(0,0,0)$
of the Wigner lattice, by a small vector $(x,y,z)$.
The energy of the system changes from the ground-state energy $E_0$ to
\begin{equation}
E(x,y,z) \mathop{\sim}_{x,y,z\to 0} E_0 + \delta E_{x,y} + \delta E_z ,
\end{equation}
where the separation of subspaces $(x,y)$ and $z$ on the harmonic level
comes from the invariance of $E(x,y,z)$ with respect to inversion
of the coordinate $z\to -z$.

The energy change $\delta E_z$ consists of three contributions,
\begin{equation}
\delta E_z = \delta E_z^{(1)} + \delta E_z^{(2)} + \delta E_z^{(3)} .
\end{equation}
The first energy change originates in direct Coulomb interactions of
the reference particle with other particles:
\begin{widetext}
\begin{equation}
\delta E_z^{(1)} =  \frac{e^2}{\varepsilon}
\sum_{j\ne 1} \left( \frac{1}{\sqrt{R_{1j}^2+z^2}} - \frac{1}{R_{1j}} \right)
\mathop{\sim}_{z\to 0} - \frac{e^2}{\varepsilon} \frac{z^2}{2} \sum_{j\ne1}
\frac{1}{R_{1j}^3} = - \frac{e^2}{\varepsilon} z^2
\frac{\sigma^{3/2}}{\sqrt{\pi}} \int_0^{\infty} {\rm d}s\, \sqrt{s}
\gamma(s) ,
\end{equation}  
where
\begin{equation}
\gamma(s) = \theta_3\left( {\rm e}^{-s\delta}\right)
\theta_3\left( {\rm e}^{-s/\delta}\right) -1 +
\theta_2\left( {\rm e}^{-s\delta}\right)
\theta_2\left( {\rm e}^{-s/\delta}\right) .
\end{equation}
The second energy change corresponds to the interaction of the reference
particle with its own images:
\begin{equation}
\delta E_z^{(2)} = \frac{e^2}{2} \left[ u_{\rm im}(0;z,z) - u_{\rm im}(0;0,0) \right]
= \frac{e^2}{2\varepsilon} \sum_{n=1}^{\infty} \Delta^{2n-1}
\left[ \frac{1}{(2n-1)d+2z} + \frac{1}{(2n-1)d-2z} - \frac{2}{(2n-1)d} \right] .
\end{equation}
Here, we do not expand the energy in small $z$ because the repulsion
of the reference particle with its self images diverges at surfaces of
the walls $z=\pm d/2$.  
The third energy change corresponds to the interaction of the reference
particle with images of other particles:
\begin{equation}
\delta E_z^{(3)} = e^2 \sum_{j\ne 1}
\left[ u_{\rm im}(R_{1j};z,0) - u_{\rm im}(R_{1j};0,0) \right] .
\end{equation}
Since
\begin{equation}
u_{\rm im}(R_{1j};z,0) - u_{\rm im}(R_{1j};0,0) = \frac{1}{\varepsilon}
\sum_{n=1}^{\infty} \Delta^n \left[ \frac{1}{\sqrt{R_{1j}^2 + (nd+z)^2}} +
\frac{1}{\sqrt{R_{1j}^2 + (nd-z)^2}}
- \frac{2}{\sqrt{R_{1j}^2 + (nd)^2}} \right] ,
\end{equation}
it holds that
\begin{equation}
\delta E_z^{(3)} \mathop{\sim}_{z\to 0} \frac{e^2}{\varepsilon} z^2
\frac{\sigma^{3/2}}{\sqrt{\pi}} \sum_{n=1}^{\infty} \Delta^n
\int_0^{\infty} {\rm d}s\, \sqrt{s} {\rm e}^{-(n\eta)^2 s}
\left( 4 n^2 \eta^2 s -2 \right) \gamma(s).
\end{equation}

The energy change along the $(x,y)$ plane $\delta E_{x,y}$ consists of
two contributions,
\begin{equation}
\delta E_{x,y} = \delta E_{x,y}^{(1)} + \delta E_{x,y}^{(2)} ,
\end{equation}
the one due to direct Coulomb interactions of the reference particle
with other particles on the monolayer
\begin{eqnarray}
\delta E_{x,y}^{(1)} & = & \frac{e^2}{\varepsilon} \left\{
\sum_{(j,k)\ne (0,0)} \left[ \frac{1}{\sqrt{(ja-x)^2+(\delta ka-y)^2}}
- \frac{1}{\sqrt{(ja)^2+(\delta ka)^2}} \right] \right. \nonumber \\ & &
\left. + \sum_{(j,k)} \left[
\frac{1}{\sqrt{\left[\left(j-\tfrac{1}{2}\right)a-x\right]^2+
\left[\delta\left(k-\tfrac{1}{2}\right)a-y\right]^2}}
-\frac{1}{\sqrt{\left[\left(j-\tfrac{1}{2}\right)a\right]^2+
\left[\delta\left(k-\tfrac{1}{2}\right)a\right]^2}} \right] \right\}
\end{eqnarray}
and the other one due to interactions of the reference particle with
electrostatic images of other particles
\begin{eqnarray}
\delta E_{x,y}^{(2)} & = & \frac{2 e^2}{\varepsilon} \sum_{n=1}^{\infty} \Delta^n
\left\{ \sum_{(j,k)\ne (0,0)} \left[
\frac{1}{\sqrt{(nd)^2+(ja-x)^2+(\delta ka-y)^2}}
- \frac{1}{\sqrt{(nd)^2+(ja)^2+(\delta ka)^2}} \right] \right.
\nonumber \\ & & \left. + \sum_{(j,k)} \left[
\frac{1}{\sqrt{(nd)^2+\left[\left(j-\tfrac{1}{2}\right)a-x\right]^2+
\left[\delta\left(k-\tfrac{1}{2}\right)a-y\right]^2}}
-\frac{1}{\sqrt{(nd)^2+\left[\left(j-\tfrac{1}{2}\right)a\right]^2+
\left[\delta\left(k-\tfrac{1}{2}\right)a\right]^2}} \right] \right\} .
\nonumber \\ & &
\end{eqnarray}  
With the aid of the substitution of variables
$x=a u$, $u\in \left( -\tfrac{1}{2},\tfrac{1}{2}\right)$ and
$y=a \delta v$, $v\in \left( -\tfrac{1}{2},\tfrac{1}{2}\right)$,
the small $u$ and $v$ expansions of these functions read as
\begin{eqnarray}
\delta E_{x,y}^{(1)} & = & \frac{e^2}{\varepsilon}\sqrt{\sigma}
\frac{4}{3\sqrt{\pi}} \int_0^{\infty} {\rm d}s\, s^{3/2}
\left[ \frac{1}{\delta^2} u^2 \alpha(s) + v^2 \beta(s) \right] \\
\delta E_{x,y}^{(2)} & = & 2 \frac{e^2}{\varepsilon}\sqrt{\sigma}
\frac{4}{3\sqrt{\pi}} \sum_{n=1}^{\infty} \Delta^n
\int_0^{\infty} {\rm d}s\, s^{3/2} {\rm e}^{-(n\eta)^2s}
\left[ \frac{1}{\delta^2} u^2 \alpha(s) + v^2 \beta(s) \right]
\end{eqnarray}
where
\begin{eqnarray}
\alpha(s) & = & \theta_3\left( {\rm e}^{-s\delta}\right)
\widetilde{\theta}_3\left( {\rm e}^{-s/\delta}\right) - \frac{\delta^2}{2}
\theta_3\left( {\rm e}^{-s/\delta}\right)
\widetilde{\theta}_3\left( {\rm e}^{-s\delta}\right) +
\theta_2\left( {\rm e}^{-s\delta}\right)
\widetilde{\theta}_2\left( {\rm e}^{-s/\delta}\right)
- \frac{\delta^2}{2} \theta_2\left( {\rm e}^{-s/\delta}\right)
\widetilde{\theta}_2\left( {\rm e}^{-s\delta}\right) , \\
\beta(s) & = & - \frac{1}{2} \theta_3\left( {\rm e}^{-s\delta}\right)
\widetilde{\theta}_3\left( {\rm e}^{-s/\delta}\right) + \delta^2
\theta_3\left( {\rm e}^{-s/\delta}\right)
\widetilde{\theta}_3\left( {\rm e}^{-s\delta}\right)
- \frac{1}{2} \theta_2\left( {\rm e}^{-s\delta}\right)
\widetilde{\theta}_2\left( {\rm e}^{-s/\delta}\right) + \delta^2
\theta_2\left( {\rm e}^{-s/\delta}\right)
\widetilde{\theta}_2\left( {\rm e}^{-s\delta}\right)
\end{eqnarray}
with the introduced functions
\begin{equation}
\widetilde{\theta}_2(q) \equiv \sum_{j=-\infty}^{\infty}
\left( j-\tfrac{1}{2} \right)^2 q^{\left( j-\tfrac{1}{2} \right)^2}
= q \frac{{\rm d}}{{\rm d}q} \theta_2(q) , \qquad
\widetilde{\theta}_3(q) \equiv \sum_{j=-\infty}^{\infty} j^2 q^{j^2}
= q \frac{{\rm d}}{{\rm d}q} \theta_3(q) .
\end{equation}  

To derive complete thermodynamics, we shift all particles $j=1,2,\ldots$
from their Wigner positions by a small vector $(x_j,y_j,z_j)$.
The total energy is a sum of the ground-state energy plus identical
local energy changes for each particle:
\begin{equation}
E\left( \{ x_j,y_j,z_j\}_{j=1}^N \right) =
E_0 + \sum_{j=1}^N \left(\delta E_{x_j,y_j} + \delta E_{z_j} \right) .
\end{equation}
The partition function is factored into
\begin{equation}
Z = {\rm e}^{-\beta F} = {\rm e}^{-\beta E_0} \left[
\int_{-a/2}^{a/2} {\rm d}x \int_{-\delta a/2}^{\delta a/2} {\rm d}y   
\int_{-d/2}^{d/2} {\rm d}z {\rm e}^{-\beta(\delta E_{x,y}+\delta E_z)} \right]^N .
\end{equation}
The pressure $P$ and the dimensionless pressure $\widetilde{P}$ are given by 
\begin{equation}
P = - 2\sigma \frac{\partial (F/N)}{\partial d} , \qquad
\widetilde{P} \equiv \frac{\beta P}{2\pi\ell_{\rm B}\sigma^2} . 
\end{equation}
In view of the previous harmonic analysis and using the substitution of
variables $t=\sqrt{\sigma} z$, the dimensionless pressure is
obtained in the form
\begin{equation} \label{press}
\widetilde{P} = \widetilde{P}_0 + \sqrt{\frac{2}{\pi\Xi}}
\frac{\partial}{\partial \eta} \ln \left[
\int_0^{\eta/2} {\rm d}t\, {\rm e}^{-f(t)} \int_0^{1/2} {\rm d}u\, {\rm e}^{-g(u)}
\int_0^{1/2} {\rm d}v\, {\rm e}^{-h(v)} \right]  
\end{equation}
where
\begin{eqnarray}
f(t) & = & \frac{1}{2} \sqrt{\frac{\Xi}{2\pi}} \sum_{n=1}^{\infty}
\Delta^{2n-1} \left[ \frac{1}{(2n-1)\eta+2t} + \frac{1}{(2n-1)\eta-2t}
- \frac{2}{(2n-1)\eta} \right] \nonumber \\ & &
+ \sqrt{\frac{\Xi}{2}} \frac{1}{\pi} t^2
\int_0^{\infty} {\rm d}s\, \sqrt{s} \left\{ -1 + \sum_{n=1}^{\infty}
\Delta^n {\rm e}^{-(n\eta)^2s} \left[ 4 (n\eta)^2 -2 \right]  
\right\} \gamma(s) , \label{ft} \\
g(u) & = & \sqrt{\frac{\Xi}{2\pi}} \frac{4}{3\sqrt{\pi}\delta^2} u^2
\int_0^{\infty} {\rm d}s\, s^{3/2} \left[ 1 + 2 \sum_{n=1}^{\infty} \Delta^n
{\rm e}^{-(n\eta)^2 s} \right] \alpha(s) ,  \\
h(v) & = & \sqrt{\frac{\Xi}{2\pi}} \frac{4}{3\sqrt{\pi}} v^2
\int_0^{\infty} {\rm d}s\, s^{3/2} \left[ 1 + 2 \sum_{n=1}^{\infty} \Delta^n
{\rm e}^{-(n\eta)^2 s} \right] \beta(s)  
\end{eqnarray}
and the ground-state pressure of phase I $\widetilde{P}_0$ is listed
in Eq. (\ref{P0eta}).

For a given dielectric jump $\Delta$, the dimensionless distance $\eta$
and the coupling constant $\Xi$, the calculation of the pressure
$\widetilde{P}$ using the symbolic language {\it Mathematica} takes
a few seconds of CPU time on PC.

The formula for the particle density is
\begin{equation} \label{rhoWSC}
\frac{\rho(z)}{2\pi\ell_{\rm B}\sigma^2} = \frac{1}{\sqrt{2\pi\Xi}}
\frac{\exp\left[-f(\sqrt{\sigma}z)\right]}{
\int_0^{\eta/2} {\rm d}t\, \exp\left[-f(t)\right]} ,  
\end{equation}
where $f(t)$ is given by (\ref{ft}).

The above procedure can be applied to phases II and III with $\bar{d}\ne 0$,
but the final formulas are too complicated to be presented.
As will be shown in Sec. \ref{sec:results}, numerical data for phases II
and III (which are dominant for intermediate and large distances $\eta$)
are very close to the exact ground-state values in the SC region.
On the other hand, in the region of small $\eta$ where phase I dominates,
the ground-state curves are no longer adequate and one has to apply
the relation (\ref{press}).
In a double-peak situation, the particle density is
calculated as the properly normalized sum of the Bolzmann factors originating
from the two peaks.  
\end{widetext}

\renewcommand{\theequation}{5.\arabic{equation}}
\setcounter{equation}{0}

\section{Monte-Carlo simulations} \label{sec:MC}

Consider two infinite uniformly charged planar surfaces with charge density $-\sigma e$ located at $z=0$ and $z=d$. The number of point counterions located between the surfaces with charge $e$ is $N_c=100$. In this region the dielectric constant is uniform, $\varepsilon$. The region outside ($z<0$ and $z>d$) is also considered as a uniform medium with dielectric constant $\varepsilon'$. The lateral size of simulation box is given by $L_{xy}=\sqrt{2\pi \Xi N_c}$ in order to keep charge neutrality. The system is replicated in $x$ and $y$ directions. The electrostatic energy of the system is
\begin{equation}
U = \frac{e^2}{2 \varepsilon}\sum_{n_x=-\infty}^{\infty}\sum_{n_y=-\infty}^{\infty}\sum_{i=1}^{N_c}\sum_{j=1}^{N_c}\frac{1}{|{\bf r}_i-{\bf r}_j+{\bf r}_{rep}|} \ ,
\end{equation}
where ${\bf r}_{i,j}$ are the positions of particles $i$ and $j$, and ${\bf r}_{rep}=n_x L_{xy}+n_y L_{xy}$ is the replication vector.
In order to calculate the electrostatic energy and consider the proper boundary conditions imposed by the dielectric contrast, we
use a modified Ewald summation together with the periodic Green functions method \cite{dosSantos16,dosSantos17}.
The equilibration is achieved after $10^6$ MC steps. The average pressures are calculated considering $10^4$ uncorrelated samples, obtained at each $10^4$ MC steps \cite{dosSantos18}.
In the next section, we turn to the comparison between the theoretical predictions and the simulation MC data.

\renewcommand{\theequation}{6.\arabic{equation}}
\setcounter{equation}{0}

\section{Comparison of MC simulations with analytic results} \label{sec:results}

\begin{widetext}
\begin{figure}[htbp]
\begin{center}
\includegraphics[clip,width=0.7\textwidth]{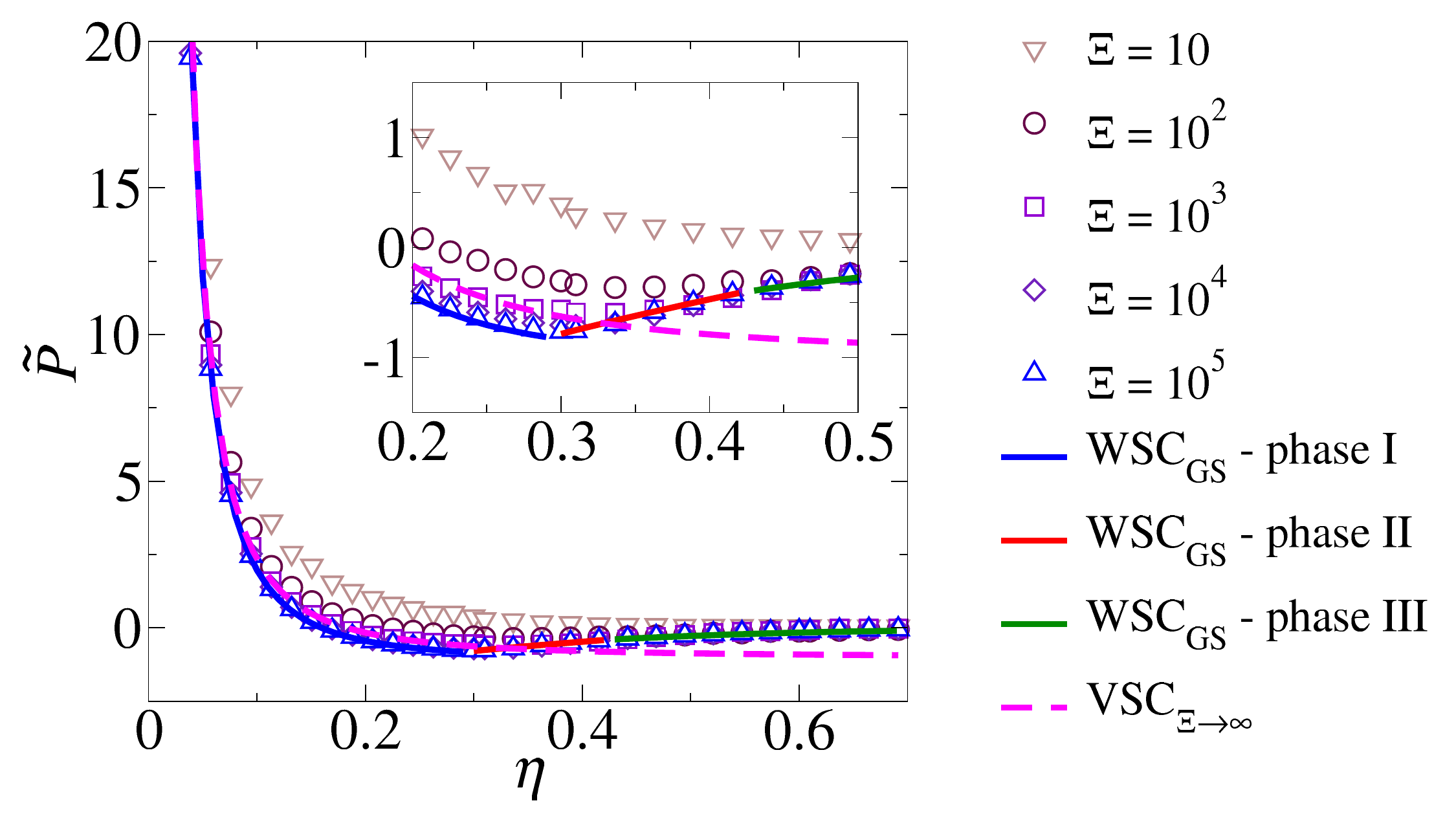}
\caption{The dimensionless pressure $\widetilde{P}$ versus the dimensionless
distance between the walls $\eta$ for the dielectric inhomogeneity
$\Delta=0.1$.
Data from MC simulations are shown by the symbols, for coupling parameter $\Xi$ varying from from 10 to $10^5$.
The ground-state pressure of the VSC theory (\ref{PVSCgs}) is displayed
by dashed curve.
The Wigner ground-state pressure is represented by solid curve;
phases I, II and III are distinguished by color. The inset shows a zoom onto the region where the ground-state pressure is minimum.}
\label{fig01A}
\end{center}
\end{figure}

The first set of figures concerns the dependence of the dimensionless
{\em ground-state} pressure $\widetilde{P}$ versus the dimensionless distance between the walls
$\eta$ for the dielectric inhomogeneity $\Delta=0.1$ (Fig. \ref{fig01A}),
$\Delta=0.5$ (Fig. \ref{fig05A}) and $\Delta=0.9$ (Fig. \ref{fig09A}).
MC simulation data are denoted by the symbol.
The ground-state pressure of the VSC theory (\ref{PVSCgs}) is displayed
by dashed curves.
The VSC curves exhibit the deficiency described in Sec. \ref{sec:VSC}
that they go to $-1$ (instead of $0$) as $\eta\to\infty$.
The Wigner ground-state pressure, calculated for phase I by using
Eq. (\ref{P0eta}) and for phases II, III by variation of formula
(\ref{phasesIII}) with respect to the bilayer distance $\bar{\eta}$
and the aspect ratio $\delta$, is represented by solid curves.
The dominance regions of phases I, II and III are distinguished by color.
The WSC curve goes correctly to $0$ for large $\eta$.

For small $\Delta=0.1$ (see Fig. \ref{fig01A}), the Wigner ground-state curve
reproduces very well MC data for strong coupling constants
$\Xi=10^3, 10^4$ and $10^5$.
The VSC ground-state curve is close to the WSC ground-state curve
in the region of small and intermediate $\eta$, up to $\eta\approx 0.3$,
and then it goes to the wrong asymptotic value
$\widetilde{P}\sim -1$ for larger $\eta$.

\begin{figure}[htbp]
\begin{center}
\includegraphics[clip,width=0.47\textwidth]{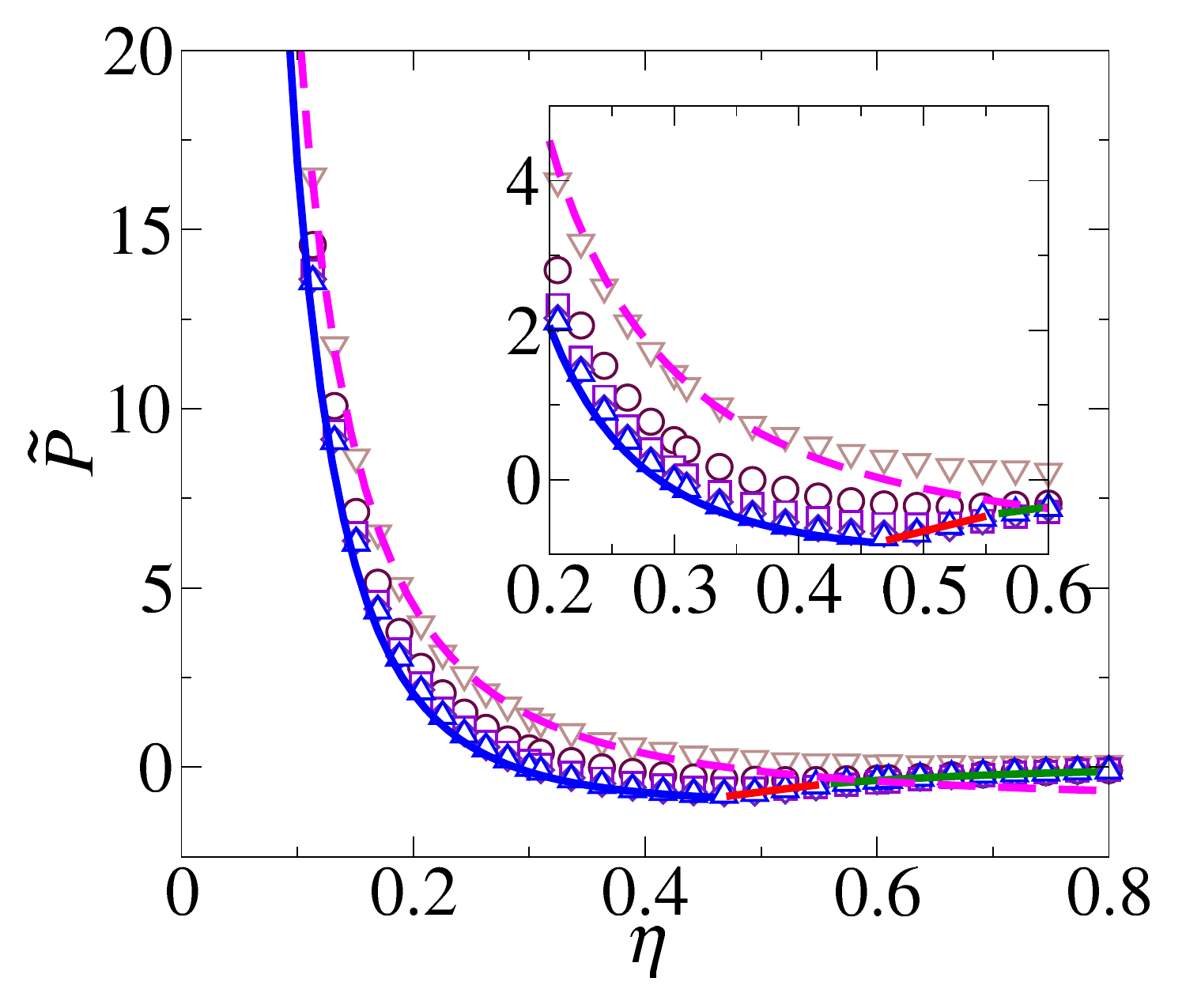}
\caption{Same as Fig. \ref{fig01A} for the dielectric inhomogeneity
$\Delta=0.5$.}
\label{fig05A}
\end{center}
\end{figure}

For intermediate $\Delta=0.5$ (see Fig. \ref{fig05A}), the Wigner ground-state
curve reproduces correctly again MC data for strong coupling constants
$\Xi=10^3, 10^4$ and $10^5$ in the whole interval of inter-wall
distances $\eta$.
The VSC ground-state curve is quite far from the WSC ground-state curve
in the region of intermediate and large values of $\eta$.

\begin{figure}[htbp]
\begin{center}
\includegraphics[clip,width=0.47\textwidth]{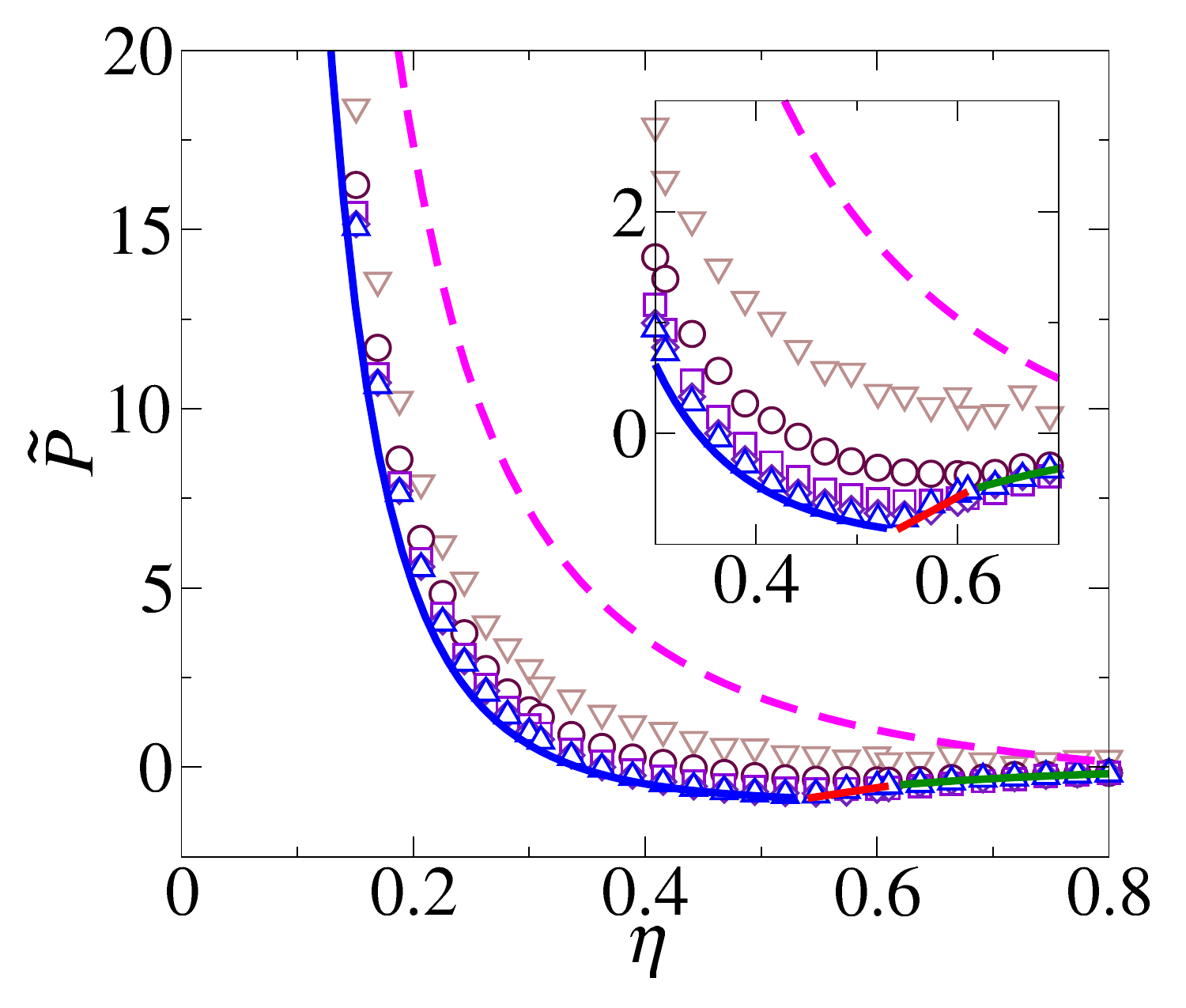}
\caption{Same as Fig. \ref{fig01A} for the dielectric inhomogeneity
$\Delta=0.9$.}
\label{fig09A}
\end{center}
\end{figure}

For large $\Delta=0.9$ (see Fig. \ref{fig09A}), the WSC ground-state curve
reproduces with a high precision MC data for strong coupling constants
$\Xi=10^3, 10^4$ and $10^5$ and reasonably for $\Xi=10$ and $10^2$,
in the whole interval of dimensionless distances $\eta$.
On the other hand, the VSC ground-state curve is significantly away from MC
data for any value of $\Xi$ and $\eta$.
The difference between the VSC and WSC ground-state curves is
larger than for previous plots.

The second set of figures concerns the dependence of the dimensionless
pressure $\widetilde{P}$ on the dimensionless distance between the walls
$\eta$ in the region of small $\eta$ where phase I dominates.
As before, we consider the dielectric inhomogeneity $\Delta=0.1$
(Fig. \ref{fig01B}), $\Delta=0.5$ (Fig. \ref{fig05B}) and
$\Delta=0.9$ (Fig. \ref{fig09B}).
MC data are denoted making use of the same symbols as in Figs \ref{fig01A},
\ref{fig05A} and \ref{fig09A}. 
The prediction of the thermal VSC theory (\ref{PVSC}), which includes
the coupling constant $\Xi$, is displayed by dashed curves.
The pressure for phase I (\ref{press}), derived within the harmonic
WSC theory, is represented by solid curves.

\begin{figure}[htbp]
\begin{center}
\includegraphics[clip,width=0.9\textwidth]{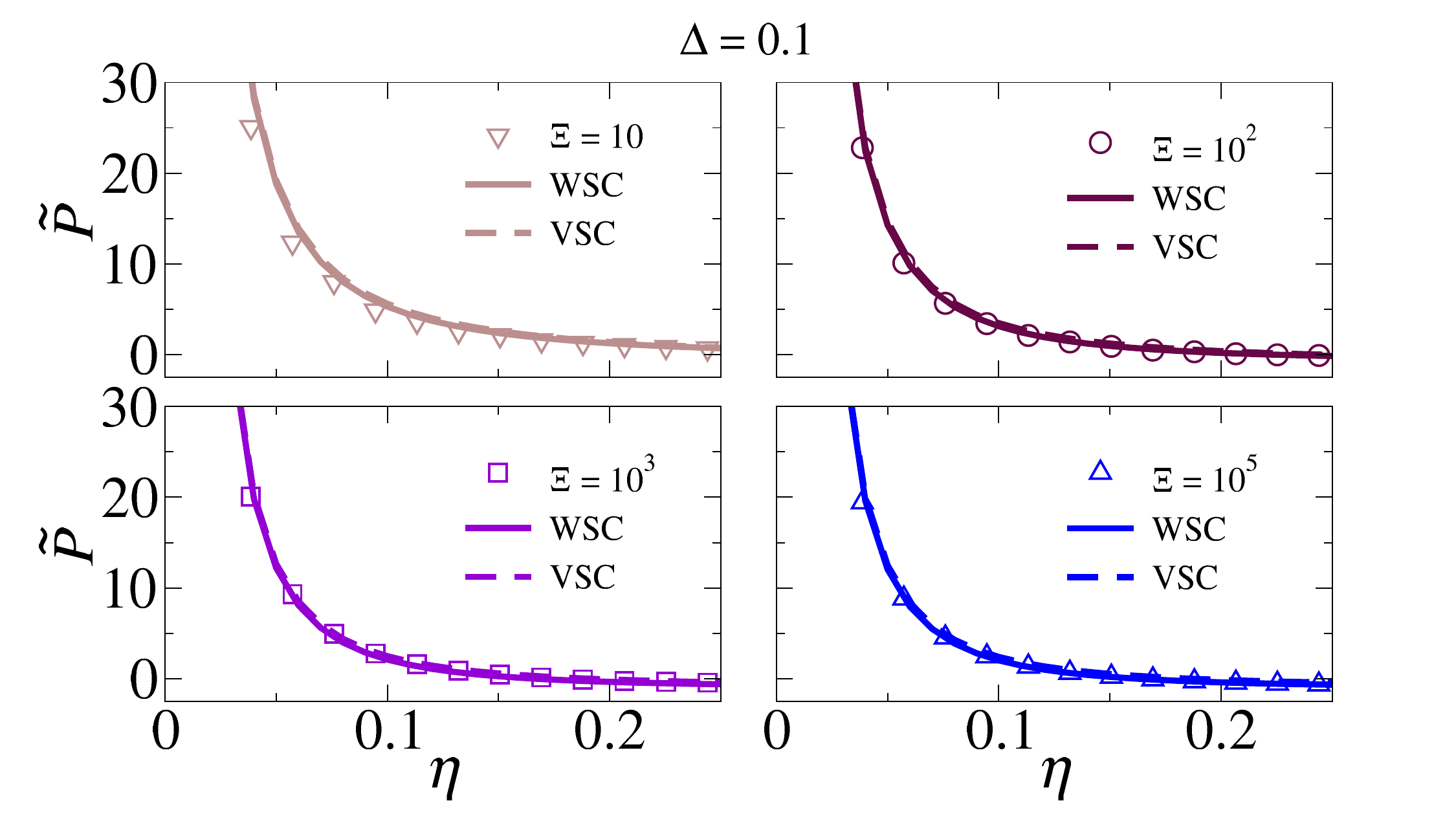}
\caption{The dimensionless pressure $\widetilde{P}$ versus the dimensionless
distance between the walls $\eta$ (constrained to the interval where
phase I dominates in the ground state) for the dielectric inhomogeneity
$\Delta=0.1$.
Data from MC simulations are denoted by upward-pointing triangles for
the coupling constant $\Xi=10^5$, by squares for $\Xi=10^3$, by circles
for $\Xi=10^2$ and by downward-pointing triangles for $\Xi=10$.
The thermal pressure of the VSC theory (\ref{PVSC}) is displayed
by dashed curves.
The thermal pressure of the WSC theory for phase I (\ref{press})
is represented by solid curves. The curves are indistinguishable at this scale.}
\label{fig01B}
\end{center}
\end{figure}

For small $\Delta=0.1$ (see Fig. \ref{fig01B}), the dashed VSC and solid WSC
curves are close to one another.
They reproduce very well MC data for all coupling constants,
including the relatively weak coupling $\Xi=10$.
This fact is remarkable because both theories were constructed on the basis of
single-particle and ground-state pictures (i.e. an approach elaborating on the $\Xi\to\infty$ limit). 

\begin{figure}[h]
\begin{center}
\includegraphics[clip,width=0.8\textwidth]{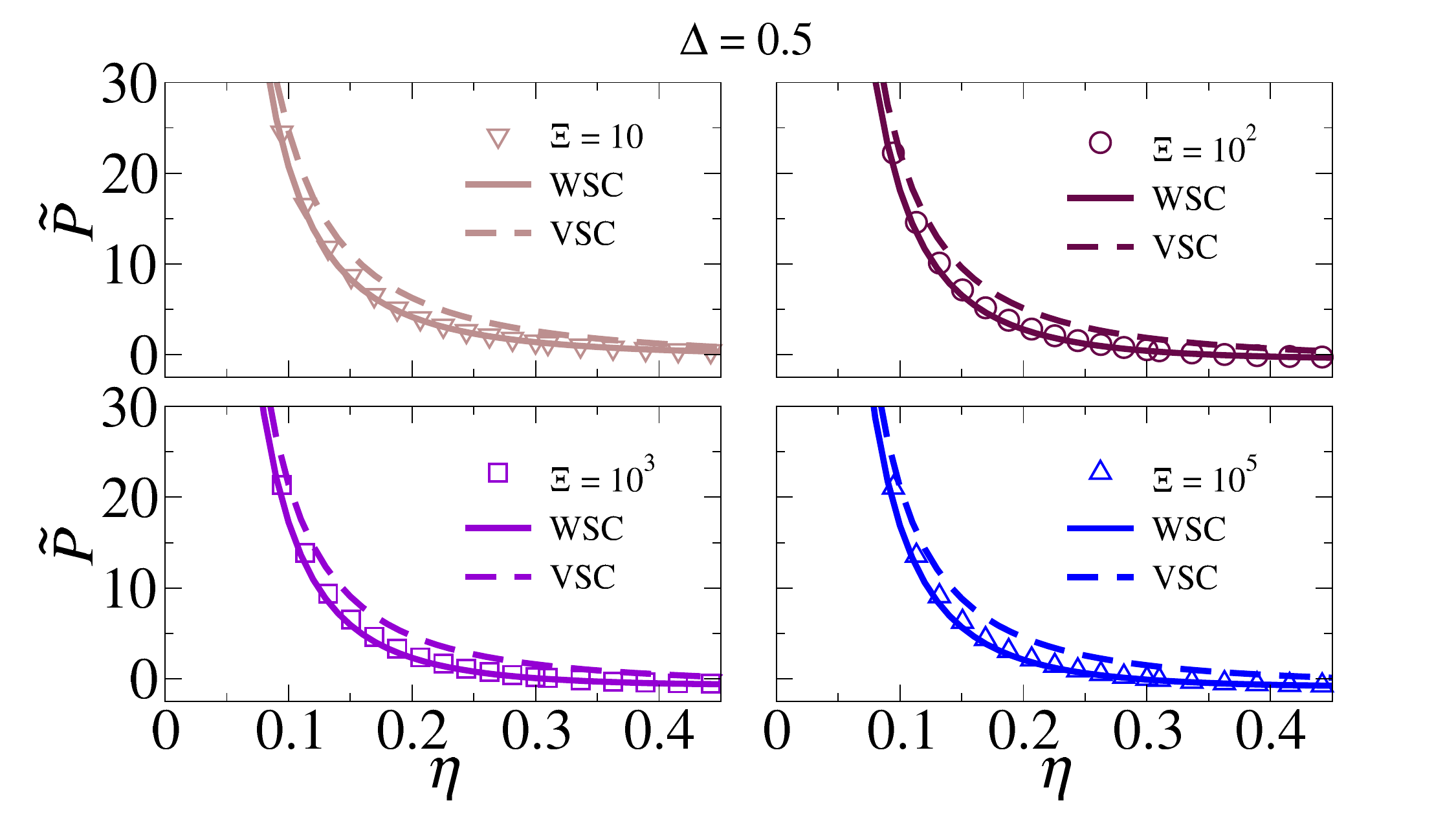}
\caption{Same as Fig. \ref{fig01B} for the dielectric inhomogeneity
$\Delta=0.5$.}
\label{fig05B}
\end{center}
\end{figure}

For intermediate $\Delta=0.5$ (see Fig. \ref{fig05B}), there is a visible
difference between the dashed VSC and solid WSC curves.
The WSC curve reproduces quite precisely MC data for all coupling constants
$\Xi=10, 10^2, 10^3$ and $10^5$, while the VSC curve is slightly above MC data.

\begin{figure}[h]
\begin{center}
\includegraphics[clip,width=0.8\textwidth]{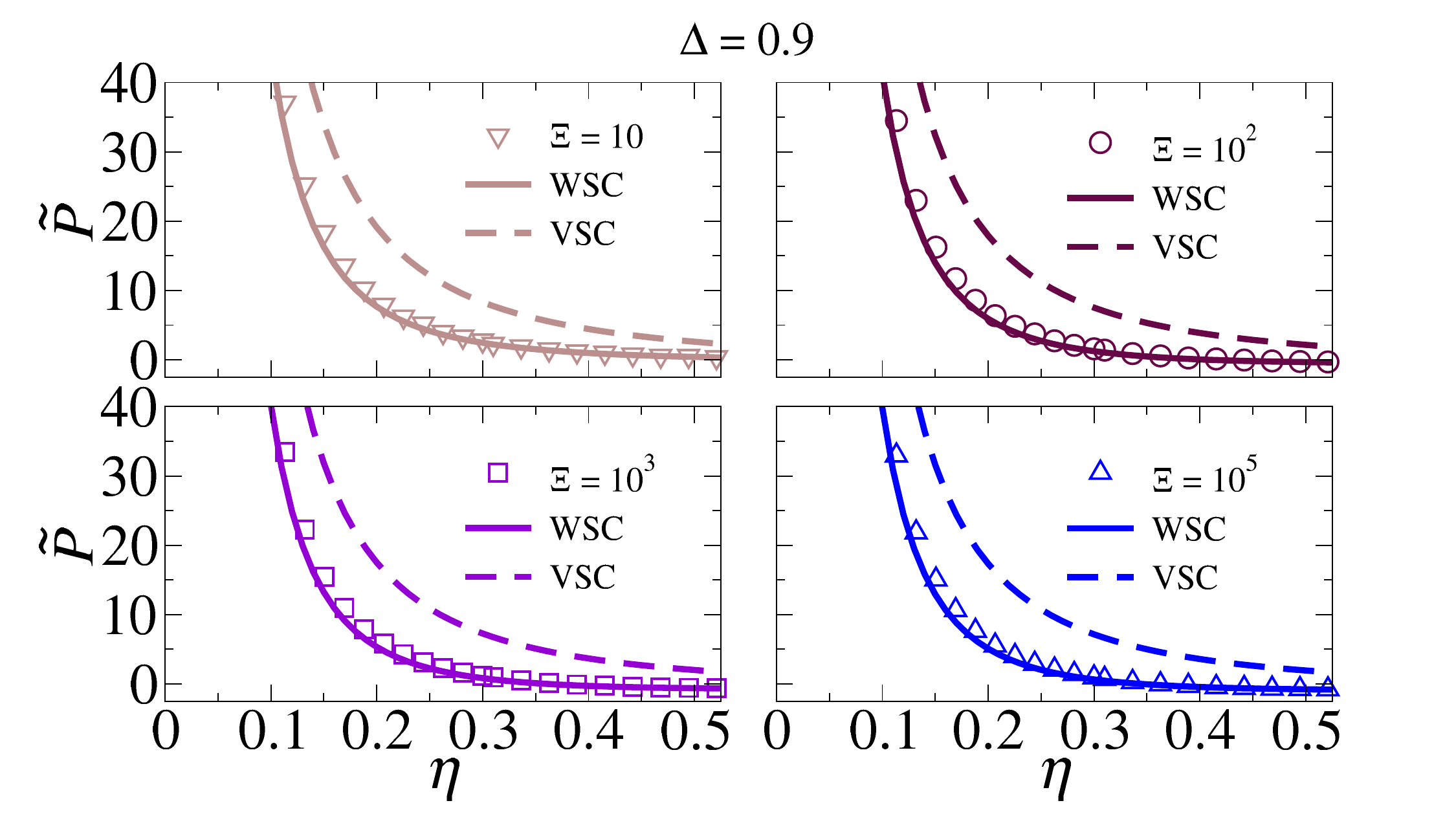}
\caption{Same as Fig. \ref{fig01B} for $\Delta=0.9$.}
\label{fig09B}
\end{center}
\end{figure}  

For large $\Delta=0.9$ (see Fig. \ref{fig09B}), the solid WSC curves
reproduce perfectly MC data for all coupling constants $\Xi$.
It is remarkable that the agreement holds down to $\Xi=10$, while the theory
stems from an infinite $\Xi$-expansion.
On the other hand, the VSC theory gives quite different pressures 
for all coupling constants, including the SC values $\Xi=10^3$ and $10^5$.
An explanation for this inadequacy may lie in the fact that at large $\Delta$,
interactions between a given ion and the images of its neighbors become more
prevalent, a many-body effect that is discarded at the level on the VSC
approach.

\begin{figure}[h]
\begin{center}
\includegraphics[width=0.58\textwidth]{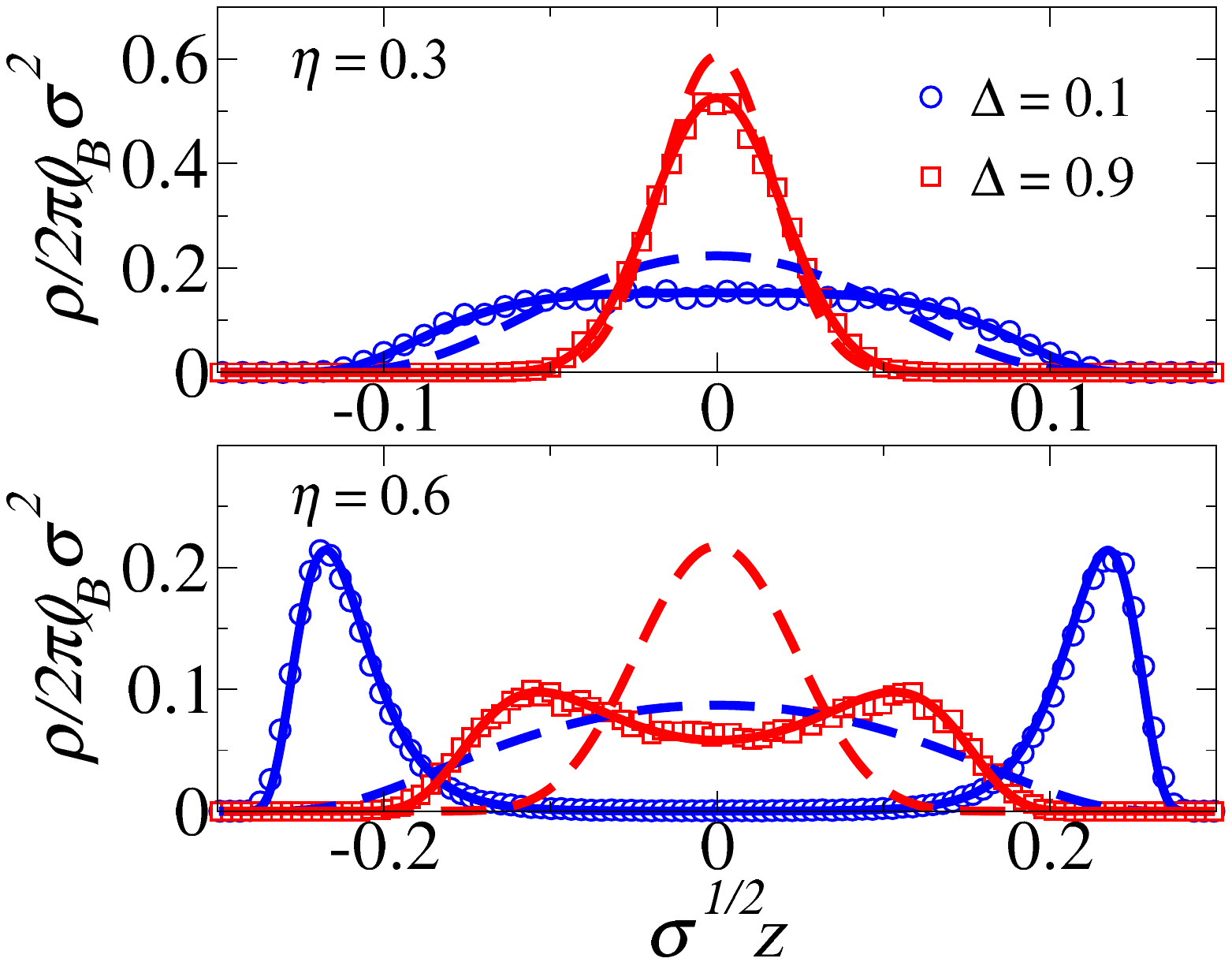}
\caption{Dependence of the dimensionless particle density
$\rho/(2\pi\ell_{\rm B}\sigma^2)$ on the dimensionless coordinate
$\sqrt{\sigma}z$ for wall distances $\eta=0.3$ (top figure)
and $\eta=0.6$ (bottom figure), for a coupling constant $\Xi=1000$.
The MC data for the dielectric jump $\Delta=0.1$ are shown by open circles
and for $\Delta=0.9$ by open squares.
The results of our WSC and VSC theories are shown by solid and
dashed curves, respectively.}
\label{densprofiles}
\end{center}
\end{figure}  

Furthermore, to illustrate the different physics at work
between the two theories, we show in Fig. \ref{densprofiles} the density
profiles for a situation where the profile is singled peaked
(top image, $\eta=0.3$), as is invariably assumed in the VSC approach,
and double-peaked (bottom image, $\eta=0.6$).
The WSC approach (solid curves) is able to switch from one regime to
the other, depending on model's parameters.
Once more, the ground state phenomenology, as investigated in \cite{Samaj12c},
allows us to predict the transition points between single- and double-peaked
profiles.
\end{widetext}  

\begin{figure}[htbp]
\begin{center}
\includegraphics[clip,width=0.52\textwidth]{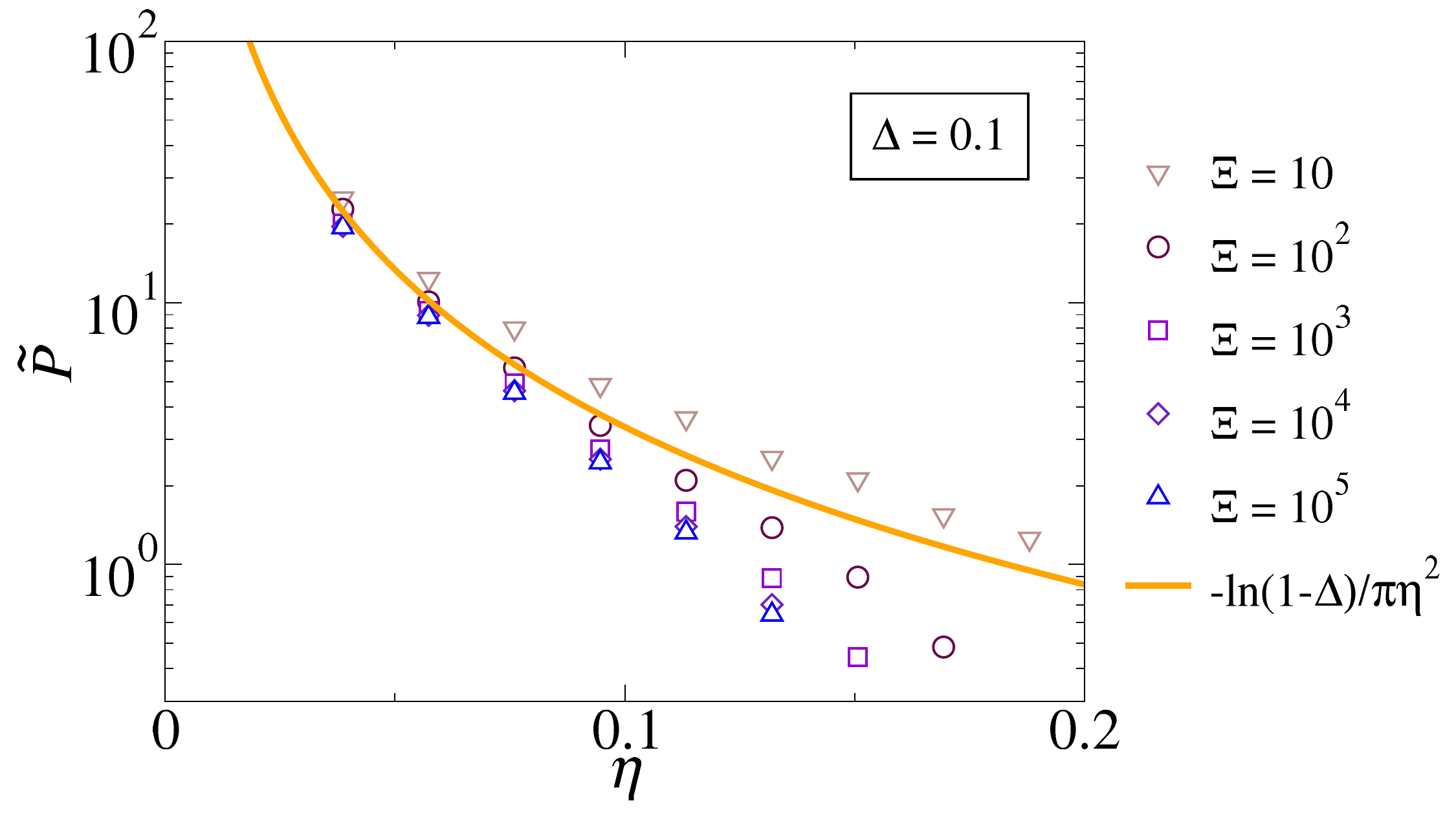}
\caption{The dimensionless pressure $\widetilde{P}$ versus the dimensionless
distance between the walls $\eta$ for the dielectric jump $\Delta=0.1$.
Comparison of MC data with the theoretical prediction of both VSC and
WSC theories $-\ln(1-\Delta)/(\pi \eta^2)$ in the asymptotic region
$\eta\to 0$ (solid curve).
Data of MC simulations are denoted by upward-pointing triangles for
the coupling constant $\Xi=10^5$, by diamonds for $\Xi=10^4$,
by squares for $\Xi=10^3$, by circles for $\Xi=10^2$ and
by downward-pointing triangles for $\Xi=10$.}
\label{fig01C}
\end{center}
\end{figure}

\begin{figure}[htbp]
\begin{center}
\includegraphics[clip,width=0.42\textwidth]{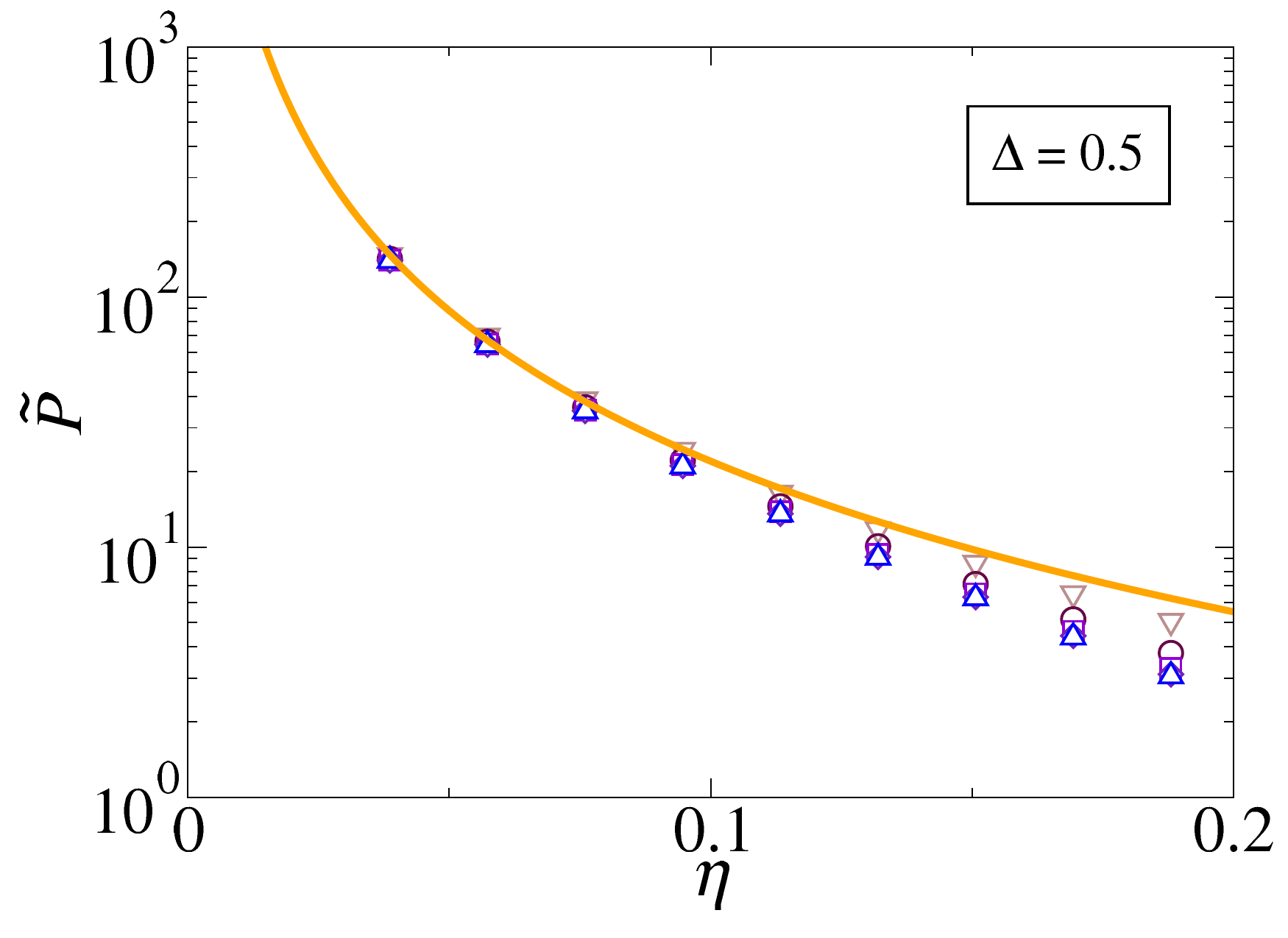}
\caption{Same as Fig. \ref{fig01C} for the dielectric inhomogeneity
$\Delta=0.5$.}
\label{fig05C}
\end{center}
\end{figure}

\begin{figure}[htbp]
\begin{center}
\includegraphics[clip,width=0.42\textwidth]{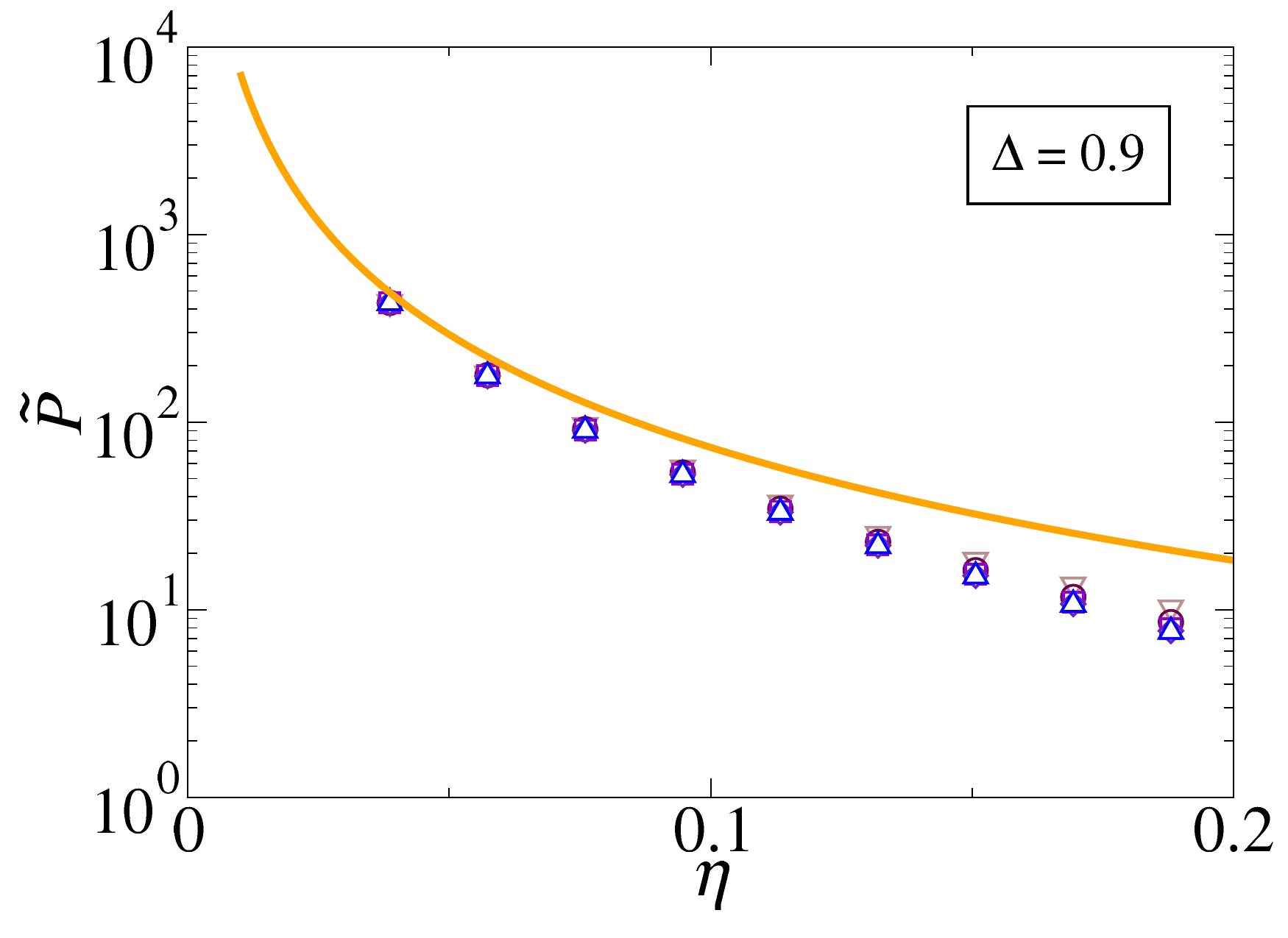}
\caption{Same as Fig. \ref{fig01C} for the dielectric inhomogeneity
$\Delta=0.9$.}
\label{fig09C}
\end{center}
\end{figure}

The third set of figures concerns the dependence of the dimensionless
pressure $\widetilde{P}$ on the dimensionless distance between the walls
$\eta$ in the region of small $\eta$.
As was shown in Sec. \ref{sec:VSC} for the VSC theory and can be shown
by analyzing the $\eta\to 0$ limit of the pressure (\ref{press}) for
the WSC theory, both theories predict that
\begin{equation} \label{prediction}
\widetilde{P} \mathop{\sim}_{\eta\to 0} - \frac{\ln(1-\Delta)}{\pi \eta^2} .  
\end{equation}
Note that this asymptotic divergence of the pressure as $\eta\to 0$
does not depend on the coupling constant $\Xi$.
We present the results for the dielectric jump $\Delta=0.1$
(Fig. \ref{fig01C}), $\Delta=0.5$ (Fig. \ref{fig05C}) and
$\Delta=0.9$ (Fig. \ref{fig09C}).
MC data are denoted by the same symbols as before.
The asymptotic prediction (\ref{prediction}) is represented by solid curves.
It is seen that for small enough $\eta$, MC data are in very good agreement with
the prediction (\ref{prediction}) for any coupling constant.

\clearpage

\renewcommand{\theequation}{6.\arabic{equation}}
\setcounter{equation}{0}

\section{Conclusion} \label{sec:conclusion}

We have studied analytically and by means of Monte Carlo simulations the problem of interacting charged plates, neutralized by mobile counterions in a solvent. The situation worked out is that when there is a dielectric mismatch between the dielectric constants of the solvent and of the medium constituting the charged plates. The strength of electrostatic interactions, as compared to thermal energy, is subsumed in the coupling parameter $\Xi$. While exact results where known in the infinite coupling limit $\Xi\to\infty$, that can be viewed as the vanishing temperature limit (hence the terminology of ``ground state'' to refer to such a situation where Wigner crystals do form), we elaborate on these results to set up a theoretical scheme that allows to investigate the coupling regime where $\Xi$ is large but not infinite. 

Once $\Xi$ is known, there are two additional dimensionless parameters that govern the physics of our problem: one related to the inter-plate distance, and the other one to the ratio of dielectric constants. 
In essence, once the ground-state has been identified for
given values of the above three parameters, our strong-coupling theory amounts to address the small fluctuations of ions' positions around their ground state locations. The signature of these fluctuations on the free energy of the global system can be computed, from which the inter-plate pressure follows by a derivation with respect to the inter-plate distance. In doing so, we obtain predictions of the Wigner-Strong Coupling type. They account for the many body nature of the SC regime, where collective ionic effects are prevalent. Neglecting these effects, as done in previous Virial Strong-Coupling approaches (VSC \cite{Kanduc07,Jho08}), may lead to poor results. Indeed, VSC theory can be viewed as stemming from an ideal-gas simplification, where a single particle physics is assumed to hold. Note that the VSC approach is already non trivial here, since due to the dielectric mismatch, a given particle interacts with an infinite number of its own dielectric images (self-images), all aligned in a line perpendicular to the plates. Due to the repulsive nature of the interactions between images, the ionic density is always maximal mid-way between the two plates, within VSC theory. While this is phenomenologically correct for small plate distances, a different piece of physics rules the range of larger distances. It is correctly captured by the WSC method.

Comparing our predictions with the Monte Carlo data indicates that our WSC approach is trustworthy not only for large $\Xi$ values, as it should, but also down to $\Xi\simeq 10$. 

Interesting perspectives are in the treatment of curved macroions, rather than planar as investigated here, together with going beyond the salt-free regime, ie accounting for mobile ions of both signs (coions and counterions). The addition of salt would significantly increase the complexity of the theory. In particular, the number of coion–counterion pairs formed~\cite{dosSantos10} as a function of concentration and coupling parameter would need to be incorporated in a future extension of the method.

\begin{acknowledgments}
L. \v{S}. is grateful to ENS de Lyon for hospitality. 
This work was supported by the Slovak Research and Development Agency under
the Contract no. APVV-24-0091 and by VEGA Grant No. 2/0089/24.
APdS acknowledges financial support from CNPq under grant 303310/2025-1
\end{acknowledgments}

\appendix

\renewcommand{\theequation}{A.\arabic{equation}}
\setcounter{equation}{0}

\section{Ground-state energy of phase I} \label{appA}
As soon as the dielectric jump $\Delta>0$ and the distance between
the plates $d$ is sufficiently small, the counterions will group into
a Wigner monolayer in the middle between the plates $(z=0)$.
They will form a hexagonal structure with a lattice spacing $a$ (in this appendix, $a$ thus has a different meaning than in section \ref{sec:VSC} where for consistency with literature it referred to $d/2$, the half distance between the plates). 
The hexagonal lattice can be represented as the union of two rectangular
lattices with sides $a$ and $\delta a$ with the aspect ratio
\begin{equation} 
\delta=\sqrt{3}
\end{equation}
shifted by half a period, see the filled and empty circles in
the first image in Fig. \ref{fig1}.
Namely, choosing site $1\equiv ({\bf R}_1,0)$ with ${\bf R}_1=(0,0)$
as the reference one, the position vectors of the first set of rectangles are
\begin{equation} \label{set1}
{\bf R}_{1j}=(ka,l\delta a) , \qquad \mbox{integer $(k,l)\ne (0,0)$.}
\end{equation}
The position vectors of the second shifted set of rectangle sites
can be numbered as
\begin{equation} \label{set2}
{\bf R}_{1j}=\left( (k-1/2)a,(l-1/2)\delta a\right) ,
\qquad \mbox{integer $(k,l)$.}
\end{equation}
The lattice spacing $a$ is determined by the requirement of the overall
neutrality of the system, namely the surface charge density $-2\sigma e$
of two walls within a rectangle with surface $\delta a^2$ must be compensated by
the charge $e$ of the particle in the center of this rectangle
plus the charge of four particles at the corners of the rectangle,
each of which contributes $e/4$ (since it is at the corners of four
adjacent rectangles), i.e.
\begin{equation} \label{defa}
-2\sigma e \delta a^2 +2 e = 0 , \qquad a=\frac{1}{\sqrt{\delta\sigma}} .
\end{equation}
In the case of the hexagonal phase I when the sub-lattices $A$ and $B$ lie
on the same surface $z=0$ (i.e. $\bar{d}=0$), the particle-particle energy
formula (\ref{part-part}) simplifies to
\begin{equation}  \label{part-part1}
\frac{E_{pp}}{N} = \frac{e^2}{2\varepsilon} \sum_{j\ne 1} \frac{1}{R_{1j}}
+ \frac{e^2}{2} \Big[ u_{\rm im}\left( 0;0,0\right) +
\sum_{j\ne 1} u_{\rm im}\left( R_{1j};0,0\right) \Big] .
\end{equation}
We are interested in the energy $E_{pp}^*/N$ regularized term by term
via neutralizing charge backgrounds.
The three terms on the rhs of (\ref{part-part1}) are analyzed successively
in what follows. 

With regard to the enumerations of sites (\ref{set1}) and (\ref{set2}),
the lattice sum in the first term on the rhs of (\ref{part-part1})
can be written as
\begin{eqnarray} 
\sum_{j\ne 1} \frac{1}{R_{1j}} & = &
\sum_{(j,k)\ne (0,0)} \frac{1}{a\sqrt{j^2+\delta^2 k^2}} \nonumber \\
& & + \sum_{(j,k)} \frac{1}{a\sqrt{\left(j-\frac{1}{2}\right)^2
+ \delta^2 \left( k-\frac{1}{2}\right)^2}} \label{coul}
\end{eqnarray}
with integer $j,k$.
Using the gamma identity
\begin{equation} \label{gammaidentity}
\frac{1}{x^{2q}} = \frac{1}{\Gamma(q)} \int_0^{\infty} {\rm d}s\, s^{q-1}
{\rm e}^{-x^2 s}  
\end{equation}  
for $q=\frac{1}{2}$, the terms in the first sum on the rhs of (\ref{coul})
can be expressed as 
\begin{eqnarray}
\frac{1}{a\sqrt{j^2+\delta^2 k^2}} & = & \frac{1}{\sqrt{\pi}a}
\int_0^{\infty} \frac{{\rm d}s}{\sqrt{s}} {\rm e}^{-(j^2+\delta^2 k^2)s}
\nonumber \\ & = & \sqrt{\frac{\sigma}{\pi}}
\int_0^{\infty} \frac{{\rm d}t}{\sqrt{t}}
{\rm e}^{-\left(\frac{j^2}{\delta}+\delta k^2\right)t} ,
\end{eqnarray}
where the substitution $t=s\delta$ and the formula (\ref{defa})
for $a$ were applied.    
Using the definition of Jacobi theta function with zero argument
\cite{Gradshteyn}
\begin{equation}
\theta_3(q) = \sum_{j=-\infty}^{\infty} q^{j^2} ,
\end{equation}
the first sum on the rhs of (\ref{coul}) can be written as
\begin{equation} \label{producttheta}
\sqrt{\frac{\sigma}{\pi}} \int_0^{\infty} \frac{{\rm d}t}{\sqrt{t}}
\left[ \theta_3\left( {\rm e}^{-t\delta}\right)
\theta_3\left( {\rm e}^{-t/\delta}\right) -1 \right] .
\end{equation}
The Poisson summation formula
\begin{equation}
\sum_{j=-\infty}^{\infty} {\rm e}^{-(j+\phi)^2 t} = \sqrt{\frac{\pi}{t}}
\sum_{j=-\infty}^{\infty} {\rm e}^{2\pi {\rm i}j\phi} {\rm e}^{-(\pi j)^2/t}  
\end{equation}
implies the asymptotic behaviors
\begin{eqnarray}
\theta_3({\rm e}^{-t}) & \displaystyle{\mathop{\sim}_{t\to\infty}} &
1 + 2 {\rm e}^{-t} + \cdots \nonumber \\
\theta_3({\rm e}^{-t}) & \displaystyle{\mathop{\sim}_{t\to 0}} &
\sqrt{\frac{\pi}{t}} \left( 1 + 2 {\rm e}^{-\pi^2/t} + \cdots \right) .
\end{eqnarray}
The product of Jacobi theta functions under integral in (\ref{producttheta})
therefore exhibits the $t\to 0$ singularity
\begin{equation} \label{singularity}
\theta_3\left( {\rm e}^{-t\delta}\right) \theta_3\left( {\rm e}^{-t/\delta}\right)
\displaystyle{\mathop{\sim}_{t\to 0}} \frac{\pi}{t} ,
\end{equation}
which makes this integral divergent.
As is shown in Ref. \cite{Samaj12b}, the effect of a neutralizing charge
background consists in subtracting from the product of Jacobi theta functions
in (\ref{producttheta}) just the singularity (\ref{singularity}), i.e.,
\begin{equation} \label{productthetareg}
\sqrt{\frac{\sigma}{\pi}} \int_0^{\infty} \frac{{\rm d}t}{\sqrt{t}}
\left[ \theta_3\left( {\rm e}^{-t\delta}\right)
\theta_3\left( {\rm e}^{-t/\delta}\right) -1 - \frac{\pi}{t} \right] .
\end{equation}
This integral is finite.
Similarly, the second sum on the rhs of (\ref{coul}) can be written as
\begin{equation} \label{producttheta2}
\sqrt{\frac{\sigma}{\pi}}
\int_0^{\infty} \frac{{\rm d}t}{\sqrt{t}}
\theta_2\left( {\rm e}^{-t\delta}\right)
\theta_2\left( {\rm e}^{-t/\delta}\right) ,
\end{equation}
where
\begin{equation}
\theta_2(q) = \sum_{j=-\infty}^{\infty} q^{\left(j-\frac{1}{2}\right)^2} 
\end{equation}
is another Jacobi theta function of zero argument \cite{Gradshteyn}
with the asymptotic behaviors
\begin{eqnarray}
\theta_2({\rm e}^{-t}) & \displaystyle{\mathop{\sim}_{t\to\infty}} &
2 {\rm e}^{-t/4} + \cdots \nonumber \\
\theta_3({\rm e}^{-t}) & \displaystyle{\mathop{\sim}_{t\to 0}} &
\sqrt{\frac{\pi}{t}} \left( 1 - 2 {\rm e}^{-\pi^2/t} + \cdots \right) .
\end{eqnarray}
As before, the effect of a neutralizing charge background
consists in subtracting from the product of Jacobi theta functions
in (\ref{producttheta2}) the $t\to 0$ singularity $\pi/t$, i.e.,
\begin{equation} \label{producttheta2reg}
\sqrt{\frac{\sigma}{\pi}} \int_0^{\infty} \frac{{\rm d}t}{\sqrt{t}}
\left[ \theta_2\left( {\rm e}^{-t\delta}\right)
\theta_2\left( {\rm e}^{-t/\delta}\right) - \frac{\pi}{t} \right] .
\end{equation}
Introducing the auxiliary functions
\begin{eqnarray}
\varphi_2(t,\delta) & \equiv & 
\theta_2\left( {\rm e}^{-t\delta}\right)
\theta_2\left( {\rm e}^{-t/\delta}\right) - \frac{\pi}{t} , \label{varphi2} \\
\varphi_3(t,\delta) & \equiv & 
\theta_3\left( {\rm e}^{-t\delta}\right)
\theta_3\left( {\rm e}^{-t/\delta}\right) -1 - \frac{\pi}{t} \label{varphi3}
\end{eqnarray}
and keeping in mind that $\delta=\sqrt{3}$, the regularized
form of the Coulomb sum $\sum_{j\ne 1} 1/R_{1j} - {\rm neutr.\, backg.}$
in (\ref{part-part1}) reads as
\begin{equation} \label{reg1}
\sqrt{\frac{\sigma}{\pi}} \int_0^{\infty} \frac{{\rm d}t}{\sqrt{t}}
\left[ \varphi_2(t,\sqrt{3}) + \varphi_3(t,\sqrt{3}) \right] .
\end{equation}  

The second term on the rhs of (\ref{part-part1}) contains
the interaction potential of the particle with its own images
$u_{\rm im}(0;0,0)$.
It is determined by formula (\ref{image}) as follows
\begin{equation} \label{reg2}
u_{\rm im}(0;0,0) = \frac{2}{\varepsilon d} \sum_{n=1}^{\infty}
\frac{\Delta^n}{n} = - \frac{2}{\varepsilon d} \ln (1-\Delta) .
\end{equation}
This potential is finite and there is no need for its regularization.

The third term on the rhs of (\ref{part-part1}) contains
the interaction potential of the reference particle $1$ at point
$({\bf R}_1,0)$ with all images of the particle at point $({\bf R}_j,0)$.
According to (\ref{image}), this potential takes the form
\begin{equation} \label{imagepot}
u_{\rm im}(R_{1j};0,0) = \frac{2}{\varepsilon} \sum_{n=1}^{\infty}
\frac{\Delta^n}{\sqrt{R_{1j}^2 + (n d)^2}} .
\end{equation}
As before, for each term in the summation we apply the integral
transformation
\begin{eqnarray}
\frac{1}{\sqrt{R_{1j}^2+(n d)^2}} = \sqrt{\frac{\sigma}{\pi}}
\int_0^{\infty} \frac{{\rm d}t}{\sqrt{t}}
{\rm e}^{-[\sigma R_{1j}^2 + (n\eta)^2]t} , \phantom{aaa} 
\end{eqnarray}
where
\begin{equation} \label{eta}
\eta = \sqrt{\sigma} d
\end{equation}
is the dimensionless distance between the walls.
Summing over lattice sites (\ref{set1}) and (\ref{set2}) and adding
neutralizing background charges results in
\begin{eqnarray} \label{reg3}
\sum_{j\ne 1} u_{\rm im}(R_{1j};0,0) & = & \frac{2}{\varepsilon}
\sqrt{\frac{\sigma}{\pi}} \int_0^{\infty} \frac{{\rm d}t}{\sqrt{t}}
\sum_{n=1}^{\infty} \Delta^n {\rm e}^{-(n\eta)^2 t} \nonumber \\
& & \times \left[ \varphi_2(t,\sqrt{3}) + \varphi_3(t,\sqrt{3}) \right] .
\end{eqnarray}

Putting together the particle-particle energy formula (\ref{part-part1})
with the regularized expressions (\ref{reg1}), (\ref{reg2}) and (\ref{reg3})
leads to the following relation for the regularized particle-particle
energy
\begin{eqnarray}
\frac{E_{pp}^*}{N (e^2/\varepsilon)\sqrt{\sigma}} & = &
- \frac{1}{\eta} \ln (1-\Delta) \nonumber \\  & & +  
\frac{1}{\sqrt{\pi}} \int_0^{\infty} \frac{{\rm d}t}{\sqrt{t}}
\left( \frac{1}{2} + \sum_{n=1}^{\infty} \Delta^n {\rm e}^{-(n\eta)^2 t} \right)
\nonumber \\ & & \times
\left[ \varphi_2(t,\sqrt{3}) + \varphi_3(t,\sqrt{3}) \right] . \label{phasesIII}
\end{eqnarray}  
The ground-state energy per particle $E_0/N$ is related to $E_{pp}^*/N$
by Eq. (\ref{gsenergy}) taken with $\bar{d}=0$, i.e.
\begin{equation}
\frac{E_0}{N} = \frac{E_{pp}^*}{N} + \frac{\pi\sigma e^2}{\varepsilon} d .
\end{equation}
Consequently,
\begin{eqnarray}
\frac{E_0}{N (e^2/\varepsilon)\sqrt{\sigma}} & = & - \frac{1}{\eta}
\ln (1-\Delta) + \pi \eta \nonumber \\ & & + 
\frac{1}{\sqrt{\pi}} \int_0^{\infty} \frac{{\rm d}t}{\sqrt{t}}
\left( \frac{1}{2} + \sum_{n=1}^{\infty} \Delta^n {\rm e}^{-(n\eta)^2 t} \right)
\nonumber \\ & & \times
\left[ \varphi_2(t,\sqrt{3}) + \varphi_3(t,\sqrt{3}) \right] . \label{E01}
\end{eqnarray}

\renewcommand{\theequation}{B.\arabic{equation}}
\setcounter{equation}{0}

\section{Ground-state energy of phases II-III} \label{appB}
For phases II and III, one applies the complete formula (\ref{part-part})
for the particle-particle energy.

The first term in the square brackets on the rhs of (\ref{part-part})
corresponds to the direct Coulomb interactions.  
For the Coulomb potential within one layer
\begin{equation}
\sum_{j\ne 1} \frac{1}{R_{1j}^{AA}} =
\sum_{(j,k)\ne (0,0)} \frac{1}{a\sqrt{j^2+\delta^2 k^2}} , 
\end{equation}  
regularized by a neutralizing background charge density, using the above
technique one gets 
\begin{equation} \label{eq1a}
\sqrt{\frac{\sigma}{\pi}} \int_0^{\infty} \frac{{\rm d}t}{\sqrt{t}}
\varphi_3(t,\delta) 
\end{equation}
where the value of the aspect ratio $\delta$ is as yet unspecified.
For the interlayer Coulomb potential
\begin{eqnarray}
\sum_{j\ne 1} \frac{1}{\sqrt{\left( R_{1j}^{AB}\right)^2 +\bar{d}^2}} =
\phantom{aaaaaaaaaaaaaaaaa}
\nonumber  \\  \sum_{(j,k)} \frac{1}{\sqrt{a^2\left(j-\frac{1}{2}\right)^2
+ \delta^2 a^2\left( k-\frac{1}{2}\right)^2 +\bar{d}^2}}
\end{eqnarray}
one gets the regularized representation 
\begin{equation} \label{eq1b}
\sqrt{\frac{\sigma}{\pi}} \int_0^{\infty} \frac{{\rm d}t}{\sqrt{t}}
{\rm e}^{-\bar{\eta}^2 t} \varphi_2(t,\delta) , 
\end{equation}
where $\bar{\eta} = \sqrt{\sigma} \bar{d}$ is the dimensionless
distance between the two particle layers.

Using (\ref{image}), the next self-image term on the rhs of (\ref{part-part})
reads as
\begin{eqnarray}
u_{\rm im}\left( 0;\frac{\bar{d}}{2},\frac{\bar{d}}{2}\right) & = &
- \frac{1}{\varepsilon d} \ln \left( 1 - \Delta^2 \right) +
\frac{1}{\varepsilon} \sum_{n=1}^{\infty} \Delta^{2n-1} \nonumber \\ & & \times
\left[ \frac{1}{(2n-1)d+\bar{d}} + \frac{1}{(2n-1)d-\bar{d}} \right] .
\nonumber \\ & & \label{eq2}
\end{eqnarray}

Without going into detail, the last image terms on the rhs
of (\ref{part-part}) can be expressed as
\begin{widetext}
\begin{equation} \label{eq3a}
\sum_{j\ne 1} u_{\rm im}\left( R_{1j}^{AA};\frac{\bar{d}}{2},\frac{\bar{d}}{2}\right)
= \frac{1}{\varepsilon} \sqrt{\frac{\sigma}{\pi}}
\int_0^{\infty} \frac{{\rm d}t}{\sqrt{t}} \varphi_3(t,\delta)
\left\{ \sum_{n=1}^{\infty} 2 \Delta^{2n} {\rm e}^{-(2n\eta)^2 t}
+ \sum_{n=1}^{\infty} \Delta^{2n-1}
\left[ {\rm e}^{-\left[(2n-1)\eta+\bar{\eta}\right]^2 t} +
{\rm e}^{-\left[(2n-1)\eta-\bar{\eta}\right]^2 t} \right] \right\} ,  
\end{equation}
\begin{equation} \label{eq3b}
\sum_j u_{\rm im}\left( R_{1j}^{AB};\frac{\bar{d}}{2},-\frac{\bar{d}}{2}\right)
= \frac{1}{\varepsilon} \sqrt{\frac{\sigma}{\pi}}
\int_0^{\infty} \frac{{\rm d}t}{\sqrt{t}} \varphi_2(t,\delta)
\left\{ \sum_{n=1}^{\infty} 2 \Delta^{2n-1} {\rm e}^{-(2n-1)^2\eta^2 t}
+ \sum_{n=1}^{\infty} \Delta^{2n}
\left[ {\rm e}^{-\left(2n\eta+\bar{\eta}\right)^2 t} +
{\rm e}^{-\left(2n\eta-\bar{\eta}\right)^2 t} \right] \right\} .
\end{equation}

Putting together equations (\ref{eq1a}), (\ref{eq1b}), (\ref{eq2}),
(\ref{eq3a}), (\ref{eq3b}) and the relation (\ref{gsenergy}),
the energy per particle is given by
\begin{eqnarray}
\frac{E(\bar{\eta},\delta)}{N (e^2/\varepsilon)\sqrt{\sigma}} & = &  
- \frac{1}{2\eta} \ln \left( 1 - \Delta^2 \right) + \frac{1}{2}
\sum_{n=1}^{\infty} \Delta^{2n-1} \left[ \frac{1}{(2n-1)\eta+\bar{\eta}}
+ \frac{1}{(2n-1)\eta-\bar{\eta}} \right] + \pi (\eta-\bar{\eta})
\nonumber \\ & & + \frac{1}{\sqrt{\pi}}
\int_0^{\infty} \frac{{\rm d}t}{\sqrt{t}} \varphi_3(t,\delta)
\left\{ \frac{1}{2} + \sum_{n=1}^{\infty} \Delta^{2n} {\rm e}^{-(2n\eta)^2 t}
+ \frac{1}{2} \sum_{n=1}^{\infty} \Delta^{2n-1}
\left[ {\rm e}^{-\left[(2n-1)\eta+\bar{\eta}\right]^2 t} +
{\rm e}^{-\left[(2n-1)\eta-\bar{\eta}\right]^2 t} \right] \right\} 
\nonumber \\ & & + \frac{1}{\sqrt{\pi}}
\int_0^{\infty} \frac{{\rm d}t}{\sqrt{t}} \varphi_2(t,\delta)
\left\{ \frac{1}{2} {\rm e}^{-\bar{\eta}^2 t} +
\sum_{n=1}^{\infty} \Delta^{2n-1} {\rm e}^{-(2n-1)^2\eta^2 t}
+ \frac{1}{2} \sum_{n=1}^{\infty} \Delta^{2n}
\left[ {\rm e}^{-\left[2n\eta+\bar{\eta}\right]^2 t} +
{\rm e}^{-\left[2n\eta-\bar{\eta}\right]^2 t} \right] \right\} .
\nonumber \\ & & \label{E02}
\end{eqnarray}
For $\bar{\eta}=0$ and $\delta=\sqrt{3}$, this relation reduces to the
one (\ref{E01}) for phase I as it should be.
\end{widetext}

\end{document}